\documentclass[11pt]{amsart}
\usepackage[dvipsnames]{xcolor}
\usepackage{hyperref, amssymb,multicol,subcaption}
\usepackage{url, tikz}

\usetikzlibrary{cd,arrows,shapes,shapes.multipart, intersections,decorations,decorations.markings,decorations.pathreplacing,fit}
	\tikzset{thick/.style={line width=.6mm}}
	\tikzset{thick2/.style={line width=.3mm}}
	\tikzstyle{vertex}=[fill=black,circle,scale=0.4]
	\tikzstyle{tensor}=[circle,thick,draw=black,fill=blue!30,minimum size=4mm]
	\tikzstyle{bigtensor}=[circle,thick,draw=black,fill=blue!30,minimum size=6mm]
	\tikzstyle{downtriangle}
	=[regular polygon,regular polygon sides=3, shape border rotate=180,thick,draw=black] 
	\tikzstyle{uptriangle}
	=[regular polygon,regular polygon sides=3, draw=black,thick] 
	\tikzstyle{smalltensor}=[rounded rectangle,thick,draw=black,fill=blue!30,minimum height = 6mm]
	\tikzstyle{hugetensor}=[rounded rectangle,thick,draw=black,minimum width=50mm,minimum height = 6mm]
	\tikzstyle{littletensor}=[circle,thick,draw=black,fill=blue!30,minimum size=3mm]
	\tikzstyle{witharrow} = [thick,decoration={
	    markings,mark=at position 0.75 with {\arrow[scale=1.5,>=stealth]{>}}},postaction={decorate}]
	\tikzstyle{shortout} = [shorten <=1.4mm,thick,decoration={
	    markings,mark=at position 0.75 with {\arrow[scale=1.5,>=stealth]{>}}},postaction={decorate}]
	\tikzstyle{shortin} = [shorten >=1.4mm,thick,decoration={
	    markings,mark=at position 0.6 with {\arrow[scale=1.5,>=stealth]{>}}},postaction={decorate}]

	\tikzstyle{longout} = [thick,shorten <=-1mm]
	\tikzstyle{longin} = [thick,shorten >=-1mm]
	\tikzstyle{underbrace} = [thick, black, decorate,decoration={brace,amplitude=10pt,mirror}]
	\tikzstyle{ub_lbl} = [black, midway,yshift=-18pt, xshift=0pt]
\tikzset{
	    bigtriangle/.style={
	        draw=black, thick,
	        shape=isosceles triangle,
	        fill=green!20,
	        minimum height=1cm,
	        minimum width=3cm,
	        shape border rotate=90,
	        isosceles triangle stretches,
	        inner sep=0pt,
	    }}

	\tikzset{
	    medtriangle/.style={
	        draw=black, thick,
	        shape=isosceles triangle,
	        fill=blue!30,
	        minimum height=.5cm,
	        minimum width=1cm,
	        shape border rotate=270,
	        isosceles triangle stretches,
	        inner sep=0pt,
	    }}

\theoremstyle{plain}
\newtheorem*{theorem*}{Theorem}

\theoremstyle{definition}

\newcommand{\<}{\langle}
\renewcommand{\>}{\rangle}
\newcommand{\ket}[1]{\vert #1 \rangle}
\newcommand{\pihat}{\widehat{\pi}}

\DeclareMathOperator{\tr}{tr}

\DeclareMathOperator{\End}{End}
\DeclareMathOperator{\parity}{parity}

\title{Modeling sequences with quantum states: a look under the hood}

\author{Tai-Danae Bradley}
\address{CUNY Graduate Center, New York, NY}
\email{tbradley@gradcenter.cuny.edu}
\author{Miles Stoudenmire}
\address{Flatiron Institute, New York, NY, A Division of the Simons Foundation}
\email{mstoudenmire@flatironinstitute.org}
\author{John Terilla}
\address{Tunnel, New York, NY}
\email{john@tunnel.tech}

\begin{document}


\begin{abstract}
Classical probability distributions on sets of sequences can be modeled using quantum states.  Here, we do so with a  quantum state that is pure and entangled.  Because it is entangled, the reduced densities that describe subsystems also carry information about the complementary subsystem. This is in contrast to the classical marginal distributions on a subsystem in which information about the complementary system has been integrated out and lost.  A training algorithm based on the density matrix renormalization group (DMRG) procedure uses the extra information contained in the reduced densities and organizes it into a tensor network model.  An understanding of the extra information contained in the reduced densities allow us to examine the mechanics of this DMRG algorithm and study the generalization error of the resulting model.  As an illustration, we work with the even-parity dataset and produce an estimate for the generalization error as a function of the fraction of the dataset used in training.
\end{abstract}

\maketitle
\tableofcontents

\section{Introduction}
In this paper, we present a deterministic algorithm for unsupervised generative modeling on strings using tensor networks.  
The algorithm is deterministic with a fixed number of steps and the resulting model has a perfect sampling algorithm that allows efficient sampling from marginal distributions, or sampling conditioned on a substring.  The algorithm is inspired by the density matrix renormalization group (DMRG) procedure \cite{schollwock,white,stoud_schwab}.  This approach, at its heart, involves only simple linear algebra which allows us to give a detailed ``under the hood'' look at the algorithm in action.  Our analysis illustrates how to interpret the trained model and how to go beyond worst case bounds on generalization errors.  We work through the algorithm with an exemplar dataset to produce a prediction for the generalization error as a function of the fraction used in training which well approximates the generalization error observed in experiments.


The machine learning problem of interest is to learn a probability distribution on a set of sequences from a finite training set of samples.  For us, an important technical and conceptual first step is to pass from \emph{Finite Sets} to \emph{Functions on Finite Sets}.  Functions on sets have more structure than sets themselves and we find that the extra structure is meaningful.  Furthermore, well-understood concepts and techniques in quantum physics give us powerful tools to exploit this extra structure without incurring significant algorithmic costs \cite{mps}.  We emphasize that it is not necessary that the datasets being modeled have any inherently quantum properties or interpretation.  The inductive bias of the model can be understood as a kind of low-rank factorization hypothesis---a point we expand upon in this paper. 

Reduced density operators play a central role in our model.  In a happy coincidence, they play the central role in both the model's theoretical inspiration and the training algorithm.  There is structure in reduced densities that inspire us to model classical probability distributions using a quantum model.  The training algorithm amounts to successively matching reduced densities, a process which leads inevitably to a tensor network model, which may be thought of as a sequence of compatible autoencoders.  We refer readers unfamiliar with tensor diagram notation to references such as \cite{tensornetwork, tensors,orus}.  

This paper also builds on investigations of tensor networks as models for machine learning tasks. Tensor networks have been demonstrated to give good results for supervised learning and regression tasks \cite{Novikov:2016,stoud_schwab,Stoudenmire:2018L,Glasser:2018,Guo:2018,Evenbly:2019,Liu:2019,Leichenauer:2019}. They have also been applied successfully to unsupervised, generative modeling \cite{Han:2018,Li:2018,Stokes:2019,Cheng:2019} including a study based on the parity dataset we use here \cite{Stokes:2019}.  This work focuses on the latter task, proposing and studying an alternative algorithm for optimizing MPS for generative modeling.  The expressivity of models like the one considered in this paper have been studied \cite{Glasser:2019}.  In this paper, we focus on understanding how our training algorithm learns to generalize.

\subsection*{Acknowledgments}  The authors thank Gabriel Drummond-Cole, Glen Evenbly, James Stokes, and Yiannis Vlassopoulos for helpful discussions, and are happy to acknowledge KITP Santa Barbara, the Flatiron Institute, and Tunnel for support and excellent working conditions.

\section{Densities and reduced densities}

For our purposes, the passage from classical to quantum can be thought of as the passage from \emph{Finite Sets} to \emph{Functions on Finite Sets}, which have a natural Hilbert space structure.  We are interested in probability distributions on finite sets.  The quantum version of a probability distribution is a density operator on a Hilbert space.  The quantum version of a marginal probability distribution is a reduced density operator.  The operation that plays the role of marginalization is the partial trace.  In our setup, the reduced densities contain more information than the marginal distributions associated to them and much of our work concerns this extra information.

Given a finite set $S$, one has the free vector space $V=\mathbb{C}^S$ consisting of complex valued functions on $S$, which is a Hilbert space with inner product 
\begin{equation*}
\<f|g\> = \sum_{s \in S}\overline{f(s)} g(s).
\end{equation*}
The free vector space comes with a natural map from $S\to \mathbb{C}^S$, which we recall in a moment.  To avoid confusion, it is helpful to use notation to distinguish between an element $s\in S$ and its image in $\mathbb{C}^S$, which is a vector.  Commonly, the vector image of $s$ is denoted with a boldface font or an overset arrow.  We like the bra and ket notation, which is better when inner products are involved.  For any $s\in S$, let $|s\>$ denote the function $S \to \mathbb{C}$ that sends $s\mapsto 1$ and $s'\mapsto 0$ for $s' \neq s$.   The set $\{|s\>\}$ is an independent, orthonormal spanning set for $V$.  If one chooses an ordering on the set $S$, say $S=\{s_1. \ldots, s_d\}$, then $|s_j\>$ is identified with the $j$-th standard basis vector in $\mathbb{C}^d$, thus defining an isometric isomorphism of $V\xrightarrow{\sim}\mathbb{C}^d$ and a ``one-hot'' encoding $S \hookrightarrow \mathbb{C}^d$.  More generally, we denote elements in $V$ by ket notation $|\psi\> \in V$.  

For any $|\psi\> \in V$, there is a linear functional in $V^*$ whose value on $|\phi\>\in V$ is the inner product $\<\psi|\phi\>$.  We denote this linear functional by the succinct bra notation $\<\psi| \in V^*$.  Every linear functional in $V^*$ is of the form $\<\psi|$ for some $|\psi\>\in V$.  We have vectors $|\psi\> \in V$ and covectors $\<\psi| \in V^*$ and the map 
\[ |\psi\> \longleftrightarrow \<\psi|
\]
defines a natural isomorphism between $V$ and $V^*$.
We have chosen to distinguish between vectors and covectors with bra and ket notation; we will not imbue upper and lower indices with any special meaning.  

When several spaces $V, W, \ldots$ are in play, some tensor product symbols are suppressed.  So, for instance, if $|\psi\> \in V$ and $|\phi\> \in W$, we will write $|\psi\> |\phi\>$, or even $|\psi \phi\>$, instead of $|\psi\> \otimes |\phi\> \in V\otimes W$.  An expression like $|\phi\>\<\psi|$ is an element of $W\otimes V^*$, naturally identified with an operator $V \to W$.  The expression $|\psi\>\<\psi|$ is an element in $\End(V)$.  Here, $\End(V)$ denotes the space of all linear operators on $V$ and in the presence of a basis is identified with $\dim(V)\times \dim(V)$ matrices.  If $|\psi\>$ is a unit vector, then the operator $|\psi\>\<\psi|$ is orthogonal projection onto $|\psi\>$: it maps $|\psi\> \mapsto |\psi\>$ and maps every vector perpendicular to $|\psi\>$ to zero. 

A \emph{density operator}, or just \emph{density} for short, is a unit-trace, positive semi-definite linear operator on a Hilbert space.  Sometimes a density is called a quantum \emph{state}.    If $S$ is a finite set and $V=\mathbb{C}^S$, then a density $\rho:V \to V$ defines a probability distribution on $S$ by defining the probability $\pi_\rho:S \to \mathbb{R}$ by the \emph{Born rule}
\begin{equation}\label{BornRule}
\pi_\rho(s)=
\<s |\rho| s\>.
\end{equation}
Going the other way, there are multiple ways to define a density $\rho:V \to V$ from a classical probability distribution $\pi$ on $S$ so that $\pi_\rho=\pi$.  One way is as a diagonal operator: $\rho_{diag}:=\sum_{s \in S}\pi(s)|s\>\<s|$.  Another way is to define 
\begin{equation}\label{rho_pi}
\rho_{\pi} = |\psi\>\<\psi| \text{ where }
|\psi\>:=\sum_{s\in S} \sqrt{\pi(s)}|s\>.
\end{equation}
There exist other densities that realize $\pi$ via the Born rule, but think of the diagonal density and projection onto $|\psi\>$ as two extremes.  The density $\rho_{\pi}$ has minimal rank and $\rho_{diag}$ has maximal rank.  In the language of quantum mechanics, a state is \emph{pure} if it has rank one and is \emph{mixed} otherwise.  The degree to which a state is mixed is measured by its von Neumann entropy, $-\tr(\rho \ln(\rho))$, which ranges from zero in the case of $\rho_{\pi}$ up to the Shannon entropy of the classical distribution $\pi$ in the case of $\rho_{diag}$.  In this paper, we always use the pure state $\rho:=\rho_{\pi}$.  To summarize, we associate to any probability distribution $\pi:S \to \mathbb{R}$ the density $\rho_{\pi}:V \to V$ defined by Equation \eqref{rho_pi}, which has the property that $\pi_{\rho_\pi} = \pi.$

If a set $S$ is a Cartesian product $S= A \times B$ then the Hilbert space $\mathbb{C}^S$ decomposes as a tensor product $\mathbb{C}^S \cong \mathbb{C}^A\otimes \mathbb{C}^B$. In this case, a density $\rho:\mathbb{C}^A \otimes \mathbb{C}^B \to \mathbb{C}^A \otimes \mathbb{C}^B$ is the quantum version of a joint probability distribution $\pi:A\times B \to \mathbb{R}$. By an operation that is analogous to marginalization, $\rho$ gives rise to two densities $\rho_A:\mathbb{C}^A \to \mathbb{C}^A$ and $\rho_B:\mathbb{C}^B \to \mathbb{C}^B$ which we refer to as \emph{reduced densities}.  We now describe this operation, which is called \emph{partial trace}.  

If $X$ and $Y$ are finite dimensional vector spaces, then $\End(X\otimes Y)$ is isomorphic to $\End(X) \otimes \End(Y)$.  Using this isomorphism, there are maps
\[
\begin{tikzcd}
    & \End(X\otimes Y) \arrow[ld, "\tr_Y"'] \arrow[rd, "\tr_X"] & \\
    \End(X) & & \End(Y)
\end{tikzcd}
\]
defined by
\[
\tr_Y(f\otimes g):= f\;\tr(g)
\text{ and }
\tr_X(f\otimes g):=g\;\tr(f)
\]
for $f\in \End(X)$ and $g\in\End(Y)$.
The maps $\tr_Y$ and $\tr_X$ are called \emph{partial traces}.   The partial trace preserves both trace and positive semi-definiteness and so the image of any density $\rho\in\End(X\otimes Y)$ under partial trace defines \emph{reduced densities} $\tr_Y\rho\in \End(X)$ and $\tr_X\rho \in \End(Y)$.

It is worth noting that while we have maps $\End(X) \otimes \End(Y) \to \End(X)$ and $\End(X) \otimes \End(Y) \to \End(Y)$, there do not exist natural maps $V\otimes W \to V$ or $V\otimes W \to W$ for arbitrary vector spaces $V$ and $W$; partial trace is special, it is defined in the case that $V$ and $W$ are endomorphism spaces.

\subsection{Reconstructing a pure state from its reduced densities}\label{sec:SVD} 
We now discuss the problem of reconstructing a pure quantum state $\rho$ on a product $X\otimes Y$ from its reduced densities $\rho_X$ and $\rho_Y$.  

\begin{figure}[h]
\begin{center}
	\begin{tikzpicture}[y=1.3cm,baseline={(current bounding box.center)}]

	\node[smalltensor, fill=yellow!10, minimum width = 14mm, minimum height = 5mm] (a) at (0,1) {};
	\node[] (L) at (-.3,0) {};
	\node[] (R) at (.3,0) {};
	\node[] (topL) at (-.3,1) {};
	\node[] (topR) at (.3,1) {};

	\draw[shortout] (topL) -- (L);
	\draw[shortout] (topR) -- (R);

	\node[] at (1.25,1) {\huge$\rightsquigarrow$};

	\node[smalltensor, fill=yellow!10, minimum width = 14mm, minimum height = 5mm] (a) at (2.5,1) {};
	\node[] (L) at (2.2,0) {};
	\node[] (R) at (2.8,0) {};
	\node[] (topL) at (2.2,1) {};
	\node[] (topR) at (2.8,1) {};

	\draw[shortin] (L) -- (topL);
	\draw[shortout] (topR) -- (R);

	\node[] at (3.75,1) {\huge$\rightsquigarrow$};

	\node[smalltensor,fill=yellow!10, minimum height = 5mm] (t) at (4.75,1) {};
	\node[] (s1) at  (4.75,2) {};
	\node[] (s2) at (4.75,0) {};

	\draw[witharrow] (s1) -- (t);
	\draw[witharrow] (t) -- (s2);

	\node[] at (5.5,1) {=};

	\node[downtriangle,minimum height = 3mm,fill=blue!30] (u) at (6.25,1.7) {};
	\node[smalltensor, minimum height = 2mm, fill = black!10] (d) at (6.25,1) {};
	\node[uptriangle,fill=green!30,minimum height = 4mm] (v) at (6.25,.3) {};
	\node[] (u1) at (6.25,2.5) {};
	\node[] (v1) at (6.25,-.5){};

	\draw[longout] (u) -- (d) ;
	\draw[longin] (d) -- (v);
	\draw[witharrow] (u1) -- (u);
	\draw[witharrow] (v) -- (v1);

	\node[] at (7.25,1) {with};

	\node[uptriangle,minimum height = 3mm,fill=blue!30] (u) at (8.5,1.25) {};
	\node[downtriangle,minimum height = 3mm,fill=blue!30] (ud) at (8.5,.75) {};
	\node[] (u1) at (8.5,2.25) {};
	\node[] (u2) at (8.5,-.25){};

	\draw[thick] (u) -- (ud);
	\draw[witharrow, longin] (u1) -- (u);
	\draw[witharrow,longout] (ud) -- (u2);

	\node[] at (9.25,1) {=};

	\draw[thick] (9.75,1.5) -- (9.75,.5) {};

	\node[] at (10.5,1) {and};

	\node[uptriangle,fill=green!30,minimum height = 3mm] (v) at (11.75,1.25) {};
	\node[downtriangle,fill=green!30,minimum height = 3mm] (vd) at (11.75,.75) {};
	\node[] (v1) at (11.75,2.25) {};
	\node[] (v2) at (11.75,-.25){};

	\draw[thick] (v) -- (vd);
	\draw[witharrow,longin] (v1) -- (v);
	\draw[witharrow,longout] (vd) -- (v2);
	
	\node[] at (12.5,1) {=};

	\draw[thick] (13.25,1.5) -- (13.25,.5) {};

	\end{tikzpicture}
\end{center}

\caption{A tensor network diagram following $|\psi\>\in X \otimes Y$ through the isomorphisms $X \otimes Y \cong X^* \otimes Y \cong \hom(X,Y)$, leading to the singular value decomposition of $M=VDU^*$ with the unitarity of $V$ and $U$.
}\label{SVD}
\end{figure}
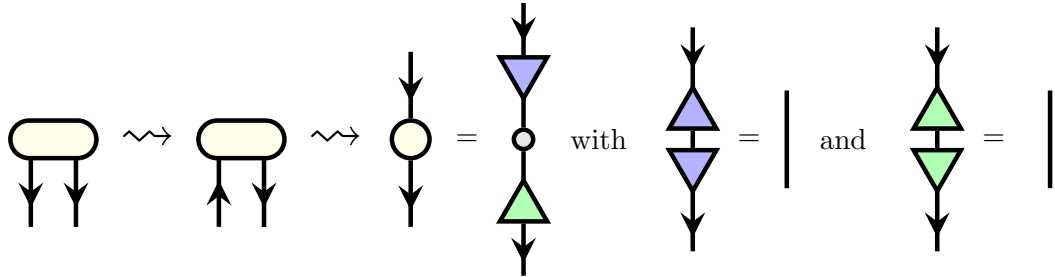

Using the isomorphism $X \cong X^*$ that is available in any finite dimensional Hilbert space,  one can view any vector $\ket{\psi}$ in a product of Hilbert spaces $X\otimes Y$ as an element of $X^*\otimes Y$, hence as a linear map $M\colon X\to Y$.  Computationally, if $\ket \psi$ is expressed using bases $\{|a\>\}$ of $X$ and $\{|b\>\}$ of $Y$ as 
\[
\ket \psi = \sum_{a,b} m_{ab} \,|a\>\otimes |b\>
\]
then the coefficients $\{m_{ab}\}$ of that sum can be reshaped into a $\dim(Y)\times \dim(X)$ matrix $M$.  A singular value decomposition (SVD) of $M$ gives a factorization $M=VDU^*$ with $V$ and $U$ unitary and $D$ diagonal as in Figure \ref{SVD}.

The columns $\{\ket{f_i}\}$ of the matrix $V$ are the left singular vectors of $M$.  They are the eigenvectors of $MM^*$ and comprise an orthonormal basis for the image of $M$.  The columns $\{\ket{e_i}\}$ of the matrix of $U$ are the right singular vectors of $M$.  They are the eigenvectors of $M^*M$, an orthonormal set of vectors spanning a subspace of $X$ isomorphic to the image of $M$.  The nonnegative real numbers $\{\sigma_i\}$ on the diagonal of $D$ are the singular values of the matrix $M$.  The matrices $M^*M$ and $MM^*$ have the same eigenvalues $\{\lambda_i\}$ which are the squares of the singular values $\lambda_i:=|\sigma_i|^2$.  The map $M$ defines a bijection between the $\{\ket{e_i}\}$ and $\{\ket{f_i}\}$.  Specifically, $M$ acts as
\begin{equation}\label{Mbijection}
\ket{e_i} \mapsto \sigma_i \ket{f_i}
\end{equation} 
and maps the perpendicular complement of the span of the $\{\ket{e_i}\}$ to zero.

\begin{figure}
\begin{center}
	\begin{tikzpicture}[y=1.3cm,baseline={(current bounding box.center)}]
	
	\node[] at (-1,0) {$\rho\:=$};

	\node[smalltensor, fill=yellow!10, minimum width = 14mm, minimum height = 5mm] (a) at (0,-0.25) {};
	\node[] (L) at (-.3,-1.25) {};
	\node[] (R) at (.3,-1.25) {};
	\node[] (topL) at (-.3,-.25) {};
	\node[] (topR) at (.3,-.25) {};

	\draw[shortout] (topL) -- (L);
	\draw[shortout] (topR) -- (R);

	\node[smalltensor, fill=yellow!10, minimum width = 14mm, minimum height = 5mm] (a) at (0,0.25) {};
	\node[] (L) at (-.3,.25) {};
	\node[] (R) at (.3,.25) {};
	\node[] (topL) at (-.3,1.25) {};
	\node[] (topR) at (.3,1.25) {};

	\draw[shortout] (L) -- (topL);
	\draw[shortout] (R) -- (topR);
		\node[] at (.75,0) {$,$};
\end{tikzpicture}
\hspace{0.5cm}
	\begin{tikzpicture}[y=1cm,baseline={(current bounding box.center)}]
	\node[] at (-1.1,1) {$\rho_X\:=$};
	\node[smalltensor, fill=yellow!10, minimum width = 14mm, minimum height = 5mm] (a) at (0,2) {};
	\node[] (L) at (-.3,1) {};
	\node[] (R) at (.3,0) {};
	\node[] (topL) at (-.3,2) {};
	\node[] (topR) at (.3,2) {};
	\draw[shortin] (L) -- (topL);
	\draw[shortin,shorten <=1.5mm,] (topR) -- (R);

	\node[smalltensor, fill=yellow!10, minimum width = 14mm, minimum height = 5mm] (a) at (0,0) {};
	\node[] (L) at (-.3,0) {};
	\node[] (topL) at (-.3,1) {};

	\draw[shortout] (L) -- (topL);
		\node[] at (.9,1) {=};

	\node[smalltensor,fill=yellow!10, minimum height = 5mm] (t1) at (1.5,1.5) {};
	\node[] (s1) at  (1.5,2.5) {};

	\node[smalltensor,fill=yellow!10, minimum height = 5mm] (t2) at (1.5,.5) {};
	\node[] (s2) at (1.5,-.5) {};

	\draw[witharrow] (s1) -- (t1);
	\draw[thick] (t1) -- (t2);
	\draw[witharrow] (t2) -- (s2);
		\node[] at (1.9,1) {$,$};
\end{tikzpicture}
\hspace{0.5cm}
	\begin{tikzpicture}[y=1cm,baseline={(current bounding box.center)}]
	\node[] at (-1.1,1) {$\rho_Y\:=$};
	\node[smalltensor, fill=yellow!10, minimum width = 14mm, minimum height = 5mm] (a) at (0,2) {};
	\node[] (L) at (-.3,0) {};
	\node[] (R) at (.3,0) {};
	\node[] (topL) at (-.3,2) {};
	\node[] (topR) at (.3,1) {};
	\draw[shortin] (L) -- (topL);
	\draw[shortout] (topR) -- (R);

	\node[smalltensor, fill=yellow!10, minimum width = 14mm, minimum height = 5mm] (a) at (0,0) {};
	\node[] (R) at (.3,1) {};
	\node[] (topR) at (.3,2) {};

	\draw[shortout] (topR) -- (R);
	\node[] at (.9,1) {=};

	\node[smalltensor,fill=yellow!10, minimum height = 5mm] (t1) at (1.5,1.5) {};
	\node[] (s1) at  (1.5,2.5) {};

	\node[smalltensor,fill=yellow!10, minimum height = 5mm] (t2) at (1.5,.5) {};
	\node[] (s2) at (1.5,-.5) {};

	\draw[witharrow] (t1) -- (s1);
	\draw[thick] (t1) -- (t2);
	\draw[witharrow] (s2) -- (t2);
			\node[] at (1.9,1) {$.$};
	\end{tikzpicture}
\end{center}
\caption{A tensor network diagram showing that \mbox{$\rho_X = M^*M$} and \mbox{$\rho_Y = MM^*$}.
}\label{MstarMisrhoX}
\end{figure}
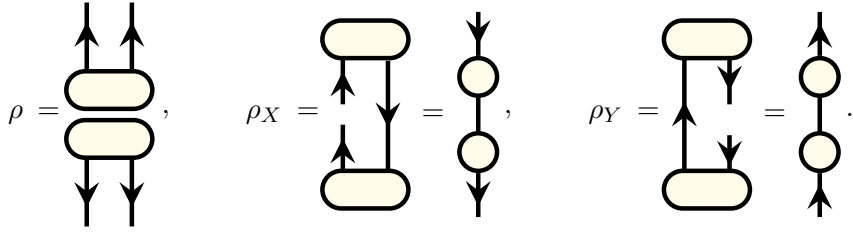
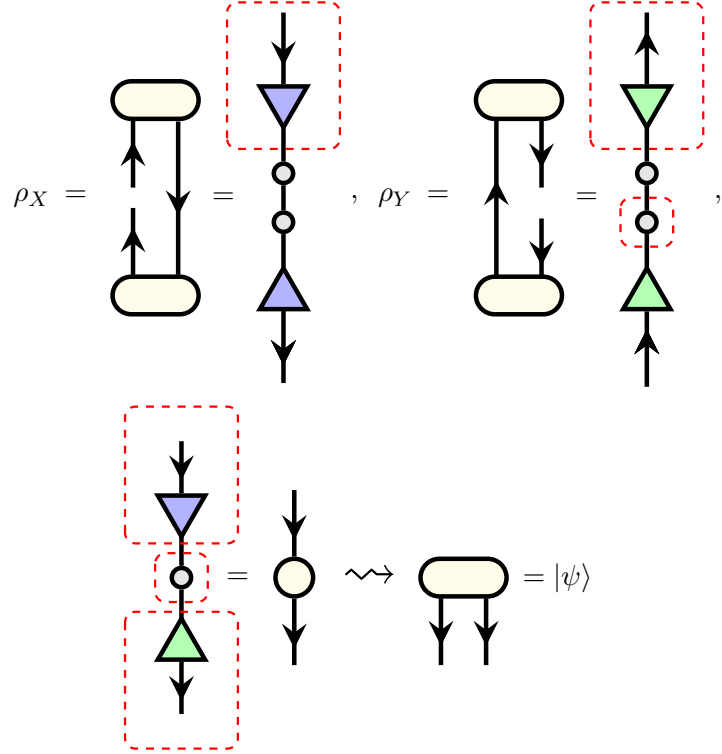
\begin{figure}[h]
\begin{center}
	\begin{tikzpicture}[y=1.3cm,baseline={(current bounding box.center)}]

	\node[] at (-3.1,0) {$\rho_X\:=$};
	\node[smalltensor, fill=yellow!10, minimum width = 14mm, minimum height = 5mm] (a) at (-1.7,1) {};
	\node[] (L) at (-2,0) {};
	\node[] (R) at (-1.4,-1) {};
	\node[] (topL) at (-2,1) {};
	\node[] (topR) at (-1.4,1) {};
	\draw[shortin] (L) -- (topL);
	\draw[shortin,shorten <=1.5mm,] (topR) -- (R);

	\node[smalltensor, fill=yellow!10, minimum width = 14mm, minimum height = 5mm] (a) at (-1.7,-1) {};
	\node[] (L) at (-2,-1) {};
	\node[] (topL) at (-2,0) {};

	\draw[shortout] (L) -- (topL);

		\node[] at (-0.8,0) {=};

	\node[downtriangle,minimum height = 3mm,fill=blue!30] (u) at (0,1) {};

	\node[smalltensor, minimum height = 2mm, fill = black!10] (mu) at (0,.25) {};	
	\node[smalltensor, minimum height = 2mm, fill = black!10] (md) at (0,-.25) {};
	\node[uptriangle,fill=blue!30,minimum height = 3mm] (d) at (0,-1) {};
	\node[] (uu) at (0,2) {};
	\node[] (dd) at (0,-2){};

	\draw[witharrow] (uu) -- (u);
	\draw[longout] (u) -- (mu) ;
	\draw[thick] (mu) -- (md);
	\draw[thick,shorten >= -.5mm] (md) -- (d);
	\draw[witharrow,shortout,shorten <=-.5mm] (d) -- (dd);

	\draw[rounded corners,thick2, dashed, red] (-.75, 0.5) rectangle (0.75, 2) {};

\end{tikzpicture}
,
\begin{tikzpicture}[y=1.3cm,baseline={(current bounding box.center)}]

	\node[] at (-3.1,0) {$\rho_Y\:=$};

	\node[smalltensor, fill=yellow!10, minimum width = 14mm, minimum height = 5mm] (a) at (-1.7,1) {};
	\node[] (L) at (-2,-1) {};
	\node[] (R) at (-1.4,-1) {};
	\node[] (topL) at (-2,1) {};
	\node[] (topR) at (-1.4,0) {};
	\draw[shortin] (L) -- (topL);
	\draw[shortout] (topR) -- (R);

	\node[smalltensor, fill=yellow!10, minimum width = 14mm, minimum height = 5mm] (a) at (-1.7,-1) {};
	\node[] (R) at (-1.4,0) {};
	\node[] (topR) at (-1.4,1) {};

	\draw[shortout] (topR) -- (R);

	\node[] at (-0.8,0) {=};

	\node[downtriangle,minimum height = 3mm,fill=green!30] (u) at (0,1) {};

	\node[smalltensor, minimum height = 2mm, fill = black!10] (mu) at (0,.25) {};	
	\node[smalltensor, minimum height = 2mm, fill = black!10] (md) at (0,-.25) {};
	\node[uptriangle,fill=green!30,minimum height = 3mm] (d) at (0,-1) {};
	\node[] (uu) at (0,2) {};
	\node[] (dd) at (0,-2){};

	\draw[witharrow] (u) -- (uu);
	\draw[longin] (mu) -- (u) ;
	\draw[thick] (md) -- (mu);
	\draw[thick,shorten <=-.5mm] (d) -- (md);
	\draw[witharrow,shortout,shorten <=-.5mm] (dd) -- (d);

	\draw[rounded corners,thick2, dashed, red] (-.75, 0.5) rectangle (0.75, 2) {};

	\draw[rounded corners,thick2, dashed, red] (-.35, 0) rectangle (0.35, -.5) {};
	\end{tikzpicture}
	,
\begin{tikzpicture}[y=1.3cm,baseline={(current bounding box.center)}]

	\node[downtriangle,minimum height = 3mm,fill=blue!30] (u) at (0.25,.7) {};
	\node[smalltensor, minimum height = 2mm, fill = black!10] (d) at (0.25,0) {};
	\node[uptriangle,fill=green!30,minimum height = 4mm] (v) at (.25,-.7) {};
	\node[] (u1) at (0.25,1.5) {};
	\node[] (v1) at (0.25,-1.5){};

	\draw[longout] (u) -- (d) ;
	\draw[longin] (d) -- (v);
	\draw[witharrow] (u1) -- (u);
	\draw[witharrow] (v) -- (v1);

	\draw[rounded corners,thick2, dashed, red] (-.5, .35) rectangle (1, 1.75) {};
	\draw[rounded corners,thick2, dashed, red] (-.5, -.35) rectangle (1, -1.75) {};

	\draw[rounded corners,thick2, dashed, red] (-.1, 0.25) rectangle (.6, -.25) {};

	\node[] at (1,0) {=};

	\node[smalltensor,fill=yellow!10, minimum height = 5mm] (t) at (1.75,0) {};
	\node[] (s1) at  (1.75,1) {};
	\node[] (s2) at (1.75,-1) {};

	\draw[witharrow] (s1) -- (t);
	\draw[witharrow] (t) -- (s2);

	\node[] at (2.75,0) {\huge$\rightsquigarrow$};

	\node[smalltensor, fill=yellow!10, minimum width = 14mm, minimum height = 5mm] (a) at (4,0) {};
	\node[] (L) at (3.7,-1) {};
	\node[] (R) at (4.3,-1) {};
	\node[] (topL) at (3.7,0) {};
	\node[] (topR) at (4.3,0) {};

	\draw[shortout] (topL) -- (L);
	\draw[shortout] (topR) -- (R);

	\node[] at (5.25,0) {$=|\psi\>$};

\end{tikzpicture}
\end{center}
\caption{Reconstructing $|\psi\>$ from the eigenvectors of $\rho_X$ and $\rho_Y$ and their shared eigenvalues.} \label{reconstructpsi}
\end{figure}

Now, given a unit vector $|\psi\> \in X\otimes Y$, we have the density $\rho = |\psi\>\<\psi| \in X \otimes Y \otimes Y^* \otimes X^*$ and the reduced densities $\rho_X: X\to X$ and $\rho_Y :Y \to Y$.  The reduced densities of $\rho$ are related to the operator $M:X \to Y$ fashioned from $|\psi\>$ as follows
\begin{equation}
\rho_X = M^*M \text{ and }\rho_Y = MM^*
\end{equation}
as illustrated in Figure \ref{MstarMisrhoX}.
The singular vectors $\{\ket{e_i}\}$ and $\{\ket{f_i}\}$ of $M$ are precisely the eigenvectors of the reduced densities.  Therefore, the density $\rho$ can be completely reconstructed from its reduced densities $\rho_X$ and $\rho_Y$.  One obtains $|\psi\>$ by gluing the eigenvectors of the reduced densities along their shared eigenvalues (Figure \ref{reconstructpsi}).  In the nondegenerate case that the eigenvalues are distinct, then there is a unique way to glue the $\{|e_i\>\}$ and the $\{|f_i\>\}$ and $|\psi\>$ is recovered perfectly.

\section{Reduced densities of classical probability distributions}
Let $\pi:S\to \mathbb{R}$ be a probability distribution and consider the density $\rho_{\pi}$ as in Equation \eqref{rho_pi}.  Suppose $S\subset A\times B$ and let $\rho_A = \tr_Y \rho$ and $\rho_B = \tr_X \rho$ denote the reduced densities where, as above, $X=\mathbb{C}^A$, $Y=\mathbb{C}^B$, and $V=X\otimes Y$.
Let us now interpret the matrix representation of these reduced densities.  We compute: 
\begin{align*}
\rho &=|\psi\>\<\psi| \\
&=\left(\sum_{(a,b)\in S}\sqrt{\pi(a,b)} |a\>\otimes |b\> \right) \otimes \left(\sum_{(a',b')\in S}\sqrt{\pi(a',b')} \<a'|\otimes \<b'| \right)\\
&=\sum_{\substack{(a,b)\in S\\(a',b')\in S}}\sqrt{\pi(a,b)}\sqrt{\pi(a',b')} \; |a\>\<a'|\otimes |b\>\<b'|
\end{align*} 
We compute the partial trace $\tr_Y(|a\>\<a'|\otimes|b\>\<b'|)=\<b|b'\> \;|a\>\<a'|$.  Since
$\<b|b'\> = 1$ if $b=b'$ and zero otherwise, we can understand the $(a,a')$ entry of the reduced density $\rho_A$ as 
\begin{equation}\label{rhox}
\left(\rho_A\right)_a^{a'} = \sum_{b\in B} \sqrt{\pi(a,b)\pi(a',b)}.
\end{equation}
In particular, the diagonal entry $\left(\rho_A\right)_a^{a}$ is $\sum_{b \in B} \pi(a,b)$ and we see the marginal distribution $\pi_A:A \to \mathbb{R}$ along the diagonal of the reduced density $\rho_A$.  We make the consistent observation that $\rho_A$ has unit trace.  The off-diagonal entries of $\rho_A$ are 
determined by the extent to which $a,a'\in A$ have the same continuations in $B$.  Note that $\rho_A$ is symmetric. 
The reduced density on $B$ is similarly given:
\begin{equation}\label{rhoy}
\left(\rho_B\right)_b^{b'} = \sum_{a\in A} \sqrt{\pi(a,b)\pi(a,b')}.
\end{equation}

So, the reduced densities of $\rho$ contains all the information of the marginal distributions $\pi_A$ and $\pi_B$ and more.  Now, let's take a look at the extra information carried by the reduced densities, which is entirely contained in the off diagonal entries.  Since the entire state, and therefore $\pi$ itself, can be reconstructed from the eigenvectors and eigenvalues of $\rho_A$ and $\rho_B$, we know that from a high level this spectral information encodes the conditional probabilities that are lost by the classical process of marginalization.  En route to decoding this spectral information, let us describe how an arbitrary density $\tau$ is a classical mixture model of pure quantum states.  If $\ket{e_1}, \ldots, \ket{e_k}$ is a basis for the image of a density $\tau$ consisting of orthonormal eigenvectors, then the corresponding eigenvalues $\lambda_1, \ldots, \lambda_k$ are nonnegative real numbers whose sum is one.  One has 
\[
\tau = \sum_{i=1}^k \lambda_i |e_i\>\<e_i|
\]
The density $\tau$ defines a probability distribution on pure states:  the probability of the pure state $|e_i\>\<e_i|$ being $\lambda_i$.  Then, $|e_i\>\<e_i|$ defines a probability distribution on the computational basis $\{s\}$ via the Born Rule:  the probability of $s$ is $\<s|e_i\>\<e_i|s\> = |\<e_i|s\>|^2$.  

We're interested in the reduced densities of $\rho=|\psi\>\<\psi|$ and in this case there exists a one-to-one correspondence $\ket{e_i} \leftrightarrow \ket{f_i}$  between eigenvectors of the reduced densities $\rho_A:=\tr_Y(\rho)$ and $\rho_B:=\tr_X(\rho)$ spanning their respective images.  
\[
\rho_A = \sum_{i=1}^k \lambda_i |e_i\>\<e_i| \text{ and }
\rho_B = \sum_{i=1}^k \lambda_i |f_i\>\<f_i|.
\]
as outlined in Section \ref{sec:SVD}.

Putting together the general picture of a density as a mixture of pure states with the reduced densities of a pure state leads one to the following paradigm.  With probability $\lambda_i$ the prefix subsystem will be in a state determined by the corresponding eigenvector $\ket{e_i}$ of $\rho_A$, and the corresponding suffix subsystem will be in a state determined by the eigenvector $\ket{f_i}$.  The vector $\ket{e_i} = \sum_a \gamma_i^a |a\>$ determines a probability distribution on the set of prefixes $A$:  the probability of the prefix $a$ is $|\gamma_i^a|^2$.  The vector $\ket{f_i} = \sum_b \beta_i^b |b\>$ determines a probability distribution on the set of suffixes $B$:  the probability of $b$ is $|\beta_i^b|^2$.

As a final remark, if we had begun with the diagonal density
\[\rho_{diag}=\sum_{(a,b)\in A\times B} \pi(a,b) \left(|a\>\otimes |b\> \right) \otimes \left(\<b|\otimes \<a| \right)
\]
whose Born distribution is also $\pi$, then
the matrices representing $\rho_A$ and $\rho_B$ would be diagonal matrices with marginal distributions on $A$ and $B$ along the diagonals and all off diagonal elements are zero.  The eigenvectors of $\rho_A$ and $\rho_B$ are simply the prefixes $\ket a$ and and suffixes $\ket b$ and carry no further information.   The process of computing reduced densities of $\rho_{diag}$ is nothing more than the process of marginalization.  We always use the pure state $\rho=|\psi\>\<\psi|$ ensuring that the reduced densities carry information about subsystem interactions.  The eigenvectors of the reduced densities, which are linear combinations of prefixes and linear combinations of suffixes, interact through their eigenvalues and capture rich information about the prefix-suffix system.

Let us summarize.  Begin with a classical probability distribution $\pi$ on a product set $S=A\times B$.  Form a density $\rho_\pi$ on $\mathbb{C}^{A \times B}$ by the formula in Equation \eqref{rho_pi}.  The reduced densities $\rho_A$ and $\rho_B$ on $\mathbb{C}^A$ and $\mathbb{C}^B$ contain marginal distributions $\pi_A$ and $\pi_B$ on their diagonals, but they are not diagonal operators.  The eigenvectors of these reduced densities encode information about prefix-suffix interactions.  The prefix-suffix interactions are tantamount to conditional probabilities and carry sufficient information to reconstruct the density $\rho$.

\subsection{Learning from samples}\label{sec:learn}
In the machine learning applications to come, the goal is to learn $\rho_\pi$ defined in Equation \eqref{rho_pi} from a set $\{s_1, \ldots, s_{N_T}\}$ of samples drawn from a probability distribution $\pi$.  Each sample $s_i$ will be a sequence $(x_1, \ldots, x_N)$ of a fixed length $N$.  The algorithm to learn the density $\rho_\pi$ on the full set of sequences $S$ is an inductive procedure.  

One only works with a density $\rho$ defined using the sample set since the density $\rho_\pi$ for the entire distribution $\pi$ is unavailable.  The procedure begins by computing the reduced density $\rho_A$ and its eigenvectors for a subsystem $A$ consisting of short prefixes.  Step by step, the size of the subsystem $A$ is increased until one reaches a point where the suffix subsystem $B$ is small.  In a final step, $\rho$ is recombined from the collected eigenvectors of $\rho_A$ for all the prefix systems $A$ and the eigenvectors and eigenvalues of $\rho_B$.  This procedure leads naturally to a tensor network approximation for $\rho$.

An important point is that the reduced density $\rho_A$ operates in a space whose dimension grows exponentially with the length of the prefix system $A$.  So, instead of computing $\rho_A$ exactly, it is computed by a sequence of approximations that keep its rank small.  The modeling hypothesis is that $\pi$ is a distribution whose corresponding quantum state $\rho_\pi$ has low rank in the sense that the reduced densities $\rho_A$ and $\rho_B$ are low rank operators for all prefix-suffix subsystems $A$ and $B$.  The large rank of the density $\rho$ witnessed from the empirical distribution drawn from $\pi$ is regarded as sampling error.  Therefore, under the modeling hypothesis, the process of replacing the empirically computed reduced densities with low rank approximations should be thought of as repairing a state damaged by sampling errors.  The low rank modeling hypothesis can lead to excellent generalization properties for the model.

Let us continue our analysis of the reduced densities as in the previous sections using notation appropriate for the machine learning algorithm.  Let $T$ be a training set of labeled samples $T=\{s_1, \ldots, s_{N_T}\}$.  We use $N_T$ for the number of training examples.  Each sample $s_i$ will be a sequence of symbols from a fixed alphabet $\Sigma$ of a fixed length $N$.  We will designate a cut to obtain a prefix $a_i$ and suffix $b_i$ whose concatenation is the sample $s_i=(a_i,b_i) \in \Sigma^N$.  This provides a decomposition of $T$ as $T\subset A\times B$ where $A=\{a_1, a_2, \ldots, a_{N_T}\}$ and $B=\{b_1, b_2, \ldots, b_{N_T}\}$ are the sampled prefixes and suffixes.
For the applications we have in mind, samples in $T$ will be distinct.  That is $(a_i,b_i) \neq (a_j,b_j)$ if $i\neq j$, though crucially it may happen that $a_i = a_j$ or $b_i=b_j$ for $i\neq j$.  Let $\pihat$ be the resulting empirical distribution on $T$ so that 
\begin{equation}
\pihat(a,b)=\begin{cases}
1/\sqrt{N_T}& \text{ if }(a,b)\in T,\\
0 &\text{otherwise.}
\end{cases}
\end{equation}
Let us look at the empirical state 
\begin{equation}\label{emp_psi}
|\psi\> = \frac{1}{\sqrt{N_T}} \sum_{i=1}^{N_T} |s_i\>,
\end{equation} the empirical density $\rho=|\psi\>\<\psi|$, and its partial trace
\begin{equation}\label{mainequation}
\rho_A = \frac{1}{N_T}\sum_{i,j=1}^{N_T} s(a_i,a_j) |a_i\>\<a_j|.
\end{equation}
Here the sum is expressed in terms of the indices $i,j$, which range over the number of samples.
The coefficient $s(a_i,a_j)$ of $|a_i\>\<a_j|$ is a nonnegative integer, namely the number of times that $a_i$ and $a_j$ have the same continuation $b_i=b_j$.  It may be convenient to have some notation for shared continuations.  For any pair $a,a'$ of elements of $A$, let $T_{a,a'}$ be the subset of $B$ consisting of shared continuations of $a$ and $a'$:
\begin{equation}\label{rhoaaprime}
T_{a,a'}=\{b\in B: (a,b) \in T \text{ and }(a',b)\in T\}.
\end{equation}
So, the $(a,a')$ entry of the matrix representing $\rho_A$ is the cardinality of the set $T_{a,a'}$ divided by an overall factor of $1/N_T$.

A similar combinatorial description holds for the reduced density on $B$,
\[\rho_B=\frac{1}{N_T}\sum_{i,j}s(b_i,b_j)|b_i\>\<b_j|\]
where $s(b_i,b_j)$ is the number of common prefixes that $b_i$ and $b_j$ share.  

The counting involved can be visualized with graphs.  Every probability distribution $\pihat$ on a Cartesian product $A\times B$ uniquely defines a weighted bipartite graph: the two vertex sets are $A$ and $B$ and the edge joining $a$ and $b$ is labeled by $\pihat(a,b).$   Here, because we assume the samples in $T$ are distinct, the graph can be simplified since $\pihat(a,b)$ is either $0$ or $1/N_T$.  We draw an edge from $a$ to $b$ if $(a,b)\in T$ and we omit the edge if $(a,b)\notin T$ and understand the probabilities to be obtained by dividing by $N_T$, which is the total number of edges in the graph.

\begin{center}
	\begin{tikzpicture}
	\node[vertex, label = left : $a_1$] (a1) at (0,.25) {};
	\node[vertex, label = left : $a_2$] (a2) at (0,-.25) {};
	\node[vertex, label = right : $b_1$] (b1) at (1,.75) {};
	\node[vertex, label = right : $b_2$] (b2) at (1,.25) {};
	\node[vertex, label = right : $b_3$] (b3) at (1,-.25) {};
	\node[vertex, label = right : $b_4$] (b4) at (1,-.75) {};

	\draw (a1) -- (b1);
	\draw (a1) -- (b4);

	\draw(a2) -- (b1);
	\draw(a2) -- (b2);
	\draw(a2) -- (b3);
	\draw(a2) -- (b4);
	\end{tikzpicture}
\end{center}

In the example above, the total number of edges is the sample size $N_T=6$.  The probability of $(a_1,b_1)=1/6$ and the probability of $(a_1,b_2)=0$.  Now we illustrate how to read off the entries of the reduced density $\rho_A$ from the graph.  There will be an overall factor of $1/N_T$ multiplied by a matrix of nonnegative integers.  The diagonal entries are $d(a)$, the degree of vertex $a$.  The $(a,a')$ entry is the number of shared suffixes, which equals the number of paths of length 2 between $a$ and $a'$, divided by $6$.

Given any graph with $|A|=2$, such as the one above, the reduced density on the prefix subsystem is equal to
\begin{equation}\label{eq:2by2}
\rho_A
=\frac{1}{N_T}\begin{bmatrix}
d_1 & s\\[5pt]
s & d_2
\end{bmatrix}
\end{equation}
where the diagonal entries are the degrees of the vertices and $s$ is the number of paths of length two, which equals the number of degree two vertices of $B$.  The denominator of the coefficient $N_T=d_1+d_2$ is the total number of edges in the graph.  The eigenvalues $\lambda_{+}$ and $\lambda_{-}$ and (unnormalized) eigenvectors $e_{+}$ and $e_{-}$ of this matrix have simple, explicit expressions in terms of the gap $G=d_1-d_2$ in the diagonal entries and the off-diagonal entry $s$.  Namely, 
\begin{equation}\label{evalues}
\lambda_{+}=\frac{N_T+\sqrt{G^2+4s^2}}{2N_T} \text{ and }
\lambda_{-}=\frac{N_T-\sqrt{G^2+4s^2}}{2N_T} 
\end{equation}
and
\begin{equation}\label{evects}
\ket{e_{+}}=
\begin{bmatrix}
\sqrt{G^2+4s^2}+ G\\
+2s
\end{bmatrix} \text{ and }
\ket{e_{-}}=
\begin{bmatrix}
\sqrt{G^2+4s^2}- G\\
-2s
\end{bmatrix}.
\end{equation}

\section{The Training Algorithm}
Suppose that $|\psi\> \in V_1 \otimes \cdots \otimes V_N$. We depict $|\psi\>$ as
\begin{center}
\begin{tikzpicture}[x=1cm,y=1.5cm,baseline={(current bounding box.center)}]
    \node[hugetensor,fill=red!10, minimum width=50mm] (v15) at (9,2) {};
        \node[] (l0) at (7,2) {};
        \node[] (l1) at (8,2) {};
        \node[] (l2) at (9,2) {};
        \node[] (l3) at (10,2) {};
        \node[] (l4) at (11,2) {};

        \node[] (n0) at (7,1) {$V_1$};
        \node[] (n1) at (8,1) {$V_{2}$};
        \node[] (n2) at (9,1) {$\cdots$};
         \node[] (n3) at (10,1) {$V_{N-1}$};
        \node[] (n4) at (11,1) {$V_{N}$};

		\draw [shortout] (l0) -- (n0);	
        \draw [shortout] (l1) -- (n1);
        \draw [shortout] (l3) -- (n3);
        \draw [shortout] (l4) -- (n4);
\end{tikzpicture}
\end{center}
There are various sorts of decompositions of such a tensor that are akin to an iterated SVD.  We will describe one decomposition that results in a factorization of $|\psi\>$ into what is called a matrix product state (MPS) or synonymously, a tensor train decomposition.  The process defines a sequence of ``bond'' spaces $\{B_k\}$ and operators $\left\{ U_k:B_{k}\otimes V_{k} \to B_{k-1}\right\}$ which can be composed $U_1 U_2 \cdots U_{N-1}U_N$ as pictured:
\begin{center}
\begin{tikzpicture}[x=1cm,y=1.5cm,baseline={(current bounding box.center)}]
	\node[bigtensor] (j0) at (0,2) {};
	\node[bigtensor] (j1) at (1,2) {};
	\node[] (j2) at (2,2) {$\cdots$};
    \node[bigtensor] (j3) at (3,2) {};
	\node[bigtensor] (j4) at (4,2) {};

    \draw[thick] (j0) -- (j1) -- (j2) -- (j3) -- (j4);
    
    \node[] (k0) at (0,1) {$V_1$};
    \node[] (k1) at (1,1) {$V_2$};
    \node[] (k3) at (3,1) {$V_{N-1}$};
    \node[] (k4) at (4,1) {$V_{N}$};
    \draw[witharrow] (j0) -- (k0);
    \draw[witharrow] (j1) -- (k1);
    \draw[witharrow] (j3) -- (k3);
    \draw[witharrow] (j4) -- (k4);

    \node[] () at (5.5,1.5) {$=$};
    \node[hugetensor,fill=red!10, minimum width=50mm] (v15) at (9,2) {};
        \node[] (l0) at (7,2) {};
        \node[] (l1) at (8,2) {};
        \node[] (l2) at (9,2) {};
        \node[] (l3) at (10,2) {};
        \node[] (l4) at (11,2) {};

        \node[] (n0) at (7,1) {$V_1$};
        \node[] (n1) at (8,1) {$V_{2}$};
        \node[] (n2) at (9,1) {$\cdots$};
         \node[] (n3) at (10,1) {$V_{N-1}$};
        \node[] (n4) at (11,1) {$V_{N}$};

		\draw [shortout] (l0) -- (n0);	
        \draw [shortout] (l1) -- (n1);
        \draw [shortout] (l3) -- (n3);
        \draw [shortout] (l4) -- (n4);

\end{tikzpicture}
\end{center}
The initial operator has form $U_1:B_1 \to V_1$ and the final tensor has the form $U_N\in B_{N-1} \otimes V_N$.  We begin with $B_1 = V_1$ and set $U_1:B_1 \to V_1$ to be the identity.  For $k=2, \ldots, N-1$ we will define $U_k$ inductively.

To describe the inductive process, first notice that for any $k=1, \ldots, N-1$, one has the tensor factorization 
\[V_1 \otimes \cdots \otimes V_N \cong \left(V_1\otimes\cdots\otimes V_k\right) \bigotimes \left(V_{k+1}\otimes\cdots \otimes V_N\right).
 \]   The operator $\alpha_k:V_1\otimes\cdots\otimes V_k \to V_{k+1}\otimes\cdots \otimes V_N$ fashioned from $|\psi\>$ may be pictured as follows:  
\begin{equation}\label{alphakpicture}
    \begin{tikzpicture}[x=1cm,y=1.5cm,baseline={(current bounding box.center)}]
        \node[] (alphak) at (-0.5,0) {$\alpha_k=$};
        \node[hugetensor,fill=red!10, minimum width=75mm] (v15) at (4,0) {};
        \node[] (j1) at (1,0) {};
        \node[] (j2) at (2,0) {};
        \node[] (j3) at (3,0) {};
        \node[] (j4) at (4,0) {};
        \node[] (j5) at (5,0) {};
        \node[] (j6) at (6,0) {};
        \node[] (j7) at (7,0) {};

        \node[] (n1) at (1,1) {$V_1$};
        \node[] (n2) at (2,1) {$V_{2}$};
        \node[] (n3) at (3,1) {$\cdots$};
        \node[] (n4) at (4,1) {$V_{k}$};
        \node[] (n5) at (5,-1) {$V_{k+1}$};

        \node[] (n6) at (6,-1) {$\cdots$};
        \node[] (n7) at (7,-1) {$V_N$};

       	\draw [shortin] (n1) -- (j1);
        \draw [shortin] (n2) -- (j2);
        \draw [shortin] (n4) -- (j4);
        \draw [shortout] (j5) -- (n5);

        \draw [shortout] (j7) -- (n7);

\end{tikzpicture}
\end{equation}
The operators $U_k$ when composed $U_1 U_2 \cdots U_k$ as below
\begin{center}
\begin{tikzpicture}[x=1cm,y=1.5cm,baseline={(current bounding box.center)}]
	\node[bigtensor] (j0) at (0,2) {};
	\node[bigtensor] (j1) at (1,2) {};
	\node[] (j2) at (2,2) {$\cdots$};
    \node[bigtensor] (j3) at (3,2) {};

    \draw[thick] (j0) -- (j1) -- (j2) -- (j3);
    \node[] (j4) at (4.75,2) {$B_{k}$};
    \node[] (k0) at (0,1) {$V_1$};
    \node[] (k1) at (1,1) {$V_2$};
    \node[] (k3) at (3,1) {$V_{k}$};
    \draw[witharrow] (j0) -- (k0);
    \draw[witharrow] (j1) -- (k1);
    \draw[witharrow] (j3) -- (k3);
    \draw[witharrow] (j4) -- (j3);
\end{tikzpicture}
\end{center}
define an operator $B_{k} \to V_1 \otimes \cdots \otimes V_k$.  One then has the composition $\beta_k:=\alpha_k U_1 U_2 \cdots U_k:B_{k} \to V_{k+1}\otimes \cdots \otimes V_N$:
\begin{center}
\begin{tikzpicture}[x=1cm,y=1.5cm,baseline={(current bounding box.center)}]
	\node[hugetensor,fill=red!10, minimum width=80mm] (v15) at (3.5,1) {};
	\node[bigtensor, label = below left : {$V_1$}] (j0) at (0,2) {};
	\node[bigtensor] (j1) at (1,2) {};
	\node[] (j2) at (2,2) {$\cdots$};
    \node[bigtensor] (j3) at (3,2) {};
    \node (j4) at (4.5,2) {};
    \node[] (j5) at (5,1) {};
    \node[] (j6) at (6,1) {};
    \node[] (j7) at (7,1) {};

    \node[] (n0) at (0,1) {};
    \node[] (n1) at (1,1) {};
    \node[] (n2) at (2,1) {};
    \node[] (n3) at (3,1) {};
    \node[] (n4) at (4,1) {};
    \node[] (j5) at (5,1) {};
    \node[] (n5) at (5,0) {$V_{k+1}$};
    \node[] (n6) at (6,0) {$\cdots$};
    \node[] (n7) at (7,0) {$V_N$};

    \draw [shortin] (j0) -- (n0);
    \draw [shortin] (j1) -- (n1);
    \draw [shortin] (j3) -- (n3);
    \draw [shortout] (j5) -- (n5);
    \draw [shortout] (j7) -- (n7);

    \draw[thick] (j0) -- (j1) -- (j2) -- (j3);
    \node[] (k5) at (4.75,2) {$B_{k}$};
    \draw[witharrow] (j4) -- (j3);
\end{tikzpicture}
\end{center}
The inductive hypothesis is that $\alpha_k U_1 U_2 \cdots U_k U_k^* \cdots U_2^* U_1^* = \alpha_k$.  Pictorally, 
\begin{center}
\begin{tikzpicture}[x=1cm,y=1.5cm,baseline={(current bounding box.center)}]
	\node[hugetensor,fill=red!10, minimum width=80mm] (v15) at (3.5,1) {};
	\node[bigtensor] (j0) at (0,2) {};
	\node[bigtensor] (j1) at (1,2) {};
	\node[] (j2) at (2,2) {$\cdots$};
    \node[bigtensor] (j3) at (3,2) {};
    \node[bigtensor,fill=blue!20] (k4) at (4,2) {};
	\node[bigtensor,fill=blue!20] (k6) at (6,2) {};
	\node[] (k5) at (5,2) {$\cdots$};
    \node[bigtensor,fill=blue!20] (k7) at (7,2) {};
    \node[] (j5) at (5,1) {};
    \node[] (j6) at (6,1) {};
    \node[] (j7) at (7,1) {};

    \node[] (n0) at (0,1) {};
    \node[] (n1) at (1,1) {};
    \node[] (n2) at (2,1) {};
    \node[] (n3) at (3,1) {};
    \node[] (n4) at (4,1) {};
    \node[] (j5) at (5,1) {};
    \node[] (n5) at (5,0) {$V_{k+1}$};
    \node[] (n6) at (6,0) {$\cdots$};
    \node[] (n7) at (7,0) {$V_N$};

    \node[] (m4) at (4,3) {$V_{k}$};
    \node[] (m5) at (5,3) {$\cdots$};
    \node[] (m6) at (6,3) {$V_2$};
    \node[] (m7) at (7,3) {$V_1$};

    \draw[witharrow] (m4) -- (k4);
    \draw[witharrow] (m6) -- (k6);
    \draw[witharrow] (m7) -- (k7);

    \draw [shortin] (j0) -- (n0);
    \draw [shortin] (j1) -- (n1);
    \draw [shortin] (j3) -- (n3);
    \draw [shortout] (j5) -- (n5);
    \draw [shortout] (j7) -- (n7);

    \draw[thick] (j0) -- (j1) -- (j2) -- (j3)-- (k4);
    \draw[thick] (k4) -- (k5) -- (k6) --(k7) ;
   
    \node[] () at (8,1.5) {$=\alpha_k$};
\end{tikzpicture}
\end{center}
In the penultimate step, one has the operator $\alpha_{N-1}U_1U_2\cdots U_{N-1}:B_{N-1} \to V_N$.  The final step is to define $U_N$ as the adjoint of this operator:  $U_N = \left(\alpha_{N-1}U_1U_2\cdots U_{N-1}\right)^*$.  
\begin{center}
\begin{tikzpicture}[x=1cm,y=1.5cm,baseline={(current bounding box.center)}]
	\node[hugetensor,fill=red!10, minimum width=80mm] (v15) at (3.5,1) {};
	\node[bigtensor] (j0) at (0,2) {};
	\node[bigtensor] (j1) at (1,2) {};
	\node[] (j3) at (3,2) {$\cdots$};
    \node[bigtensor] (j5) at (5,2) {};
	\node[bigtensor] (j6) at (6,2) {};
    \node[] (j7) at (7,1) {};

    \node[] (n0) at (0,1) {};
    \node[] (n1) at (1,1) {};
    \node[] (n2) at (2,1) {};
    \node[] (n3) at (3,1) {};
    \node[] (n4) at (4,1) {};
    \node[] (n5) at (5,1) {};
    \node[] (n6) at (6,1) {};
    \node[] (n7) at (7,0) {$V_N$};
    \node[] (k7) at (7.75,2) {$B_{N-1}$};

    \draw[witharrow] (k7) -- (j6);

    \draw [shortin] (j0) -- (n0);
    \draw [shortin] (j1) -- (n1);
    \draw [shortin] (j5) -- (n5);
    \draw [shortin] (j6) -- (n6);
    \draw [shortout] (j7) -- (n7);

    \draw[thick] (j0) -- (j1) -- (j3) -- (j5)-- (j6);
 
    \node[] () at (8,1.5) {$=$};
    \node[bigtensor] (un) at (9,1.5) {};
    \node[] (ub) at (9,.5) {$V_N$};
    \node[] (ur) at (10.75,1.5) {$B_{N-1}$};
    \draw[witharrow] (ur) --(un);
    \draw[witharrow] (un) --(ub);
    
\end{tikzpicture}
\end{center}
Therefore, the entire composition reduces nicely:
\begin{align*}
U_1 U_2 \cdots U_{N-1} U_N &= U_1 U_2 \cdots U_{N-1} U_{N-1}^* \cdots U_2^* U_1^* \alpha_{N-1}^* \\
&= \alpha_{N-1}^*
\end{align*}
The final equality follows from the adjoint of the inductive hypothesis.
The outcome $\alpha_{N-1}^*: V_N^* \to V_1 \otimes \cdots \otimes  V_{N-1}$, after a minor reshaping, is the same as $|\psi\>$.

To define the inducive step, assume the spaces $B_1, \ldots, B_{k-1}$ and operators $U_k$ have been defined and satisfy the inductive hypothesis.  Reshape the operator $B_{k-1} \to V_k \otimes V_{k+1}\otimes \cdots \otimes V_N$ as a map 
\[B_{k-1} \otimes V_k \to V_{k+1}\otimes \cdots \otimes V_N\]%
An SVD decomposition of this map yields $\alpha_{k-1} U_1 \cdots U_{k-1} = W_k D_k U_k^*$.  
\begin{center}
\begin{tikzpicture}[x=1cm,y=1.5cm,baseline={(current bounding box.center)}]
	\node[hugetensor,fill=red!10, minimum width=90mm] (v15) at (4,1) {};
	\node[bigtensor, label = below left : {$V_1$}] (j0) at (0,2) {};
	\node[bigtensor] (j1) at (1,2) {};
	\node[] (j2) at (2,2) {$\cdots$};
    \node[bigtensor] (j3) at (3,2) {};
    \node (j4) at (4.5,2) {};
    \node[] (j5) at (5.5,2) {};
    \node[] (j6) at (6,1) {};
    \node[] (j7) at (7,1) {};
    \node[] (j8) at (8,1) {};

    \node[] (n0) at (0,1) {};
    \node[] (n1) at (1,1) {};
    \node[] (n2) at (2,1) {};
    \node[] (n3) at (3,1) {};
    \node[] (n4) at (4,1) {};
    \node[] (m5) at (5,2) {$V_{k}$};
    \node[] (n5) at (5,1) {};
    \node[] (n6) at (6,0) {$V_{k+1}$};
    \node[] (n7) at (7,0) {$\cdots$};
    \node[] (n8) at (8,0) {$V_N$};

    \draw [shortin] (j0) -- (n0);
    \draw [shortin] (j1) -- (n1);
    \draw [shortin] (j3) -- (n3);
    \draw [shortin] (m5) -- (n5);
    \draw [shortout] (j6) -- (n6);
    \draw [shortout] (j8) -- (n8);

    \draw[thick] (j0) -- (j1) -- (j2) -- (j3);
    \node[label = above left : {$B_{k-1}$}] (k5) at (4.5,2) {};
    \draw[witharrow] (j4) -- (j3);

    \node at (9.5,1) {$=$};

	\node[medtriangle,minimum height = 8mm] (u) at (11,2) {};
	\node[smalltensor, minimum height = 2mm, fill = black!10] (d) at (11,1.25) {};
	\node[bigtriangle] (v) at (11,.5) {};

	\node (a1) at (10.75,2) {};
	\node (a2) at (11.25,2) {};
	\node (a3) at (10,.4) {};
	\node (a4) at (12,.4) {};

	\node[label = left : $B_{k-1}$] (b1) at (10.75,3) {};
	\node[label = right : $V_k$] (b2) at (11.25,3) {};
	\node (b3) at (10,-.75) {$V_{k+1}$};
	\node (b4) at (12,-.75) {$V_N$};

	\node at (11,-.25) {$\cdots$};

	\draw[shortin] (b1) -- (a1);
	\draw[shortin] (b2) -- (a2);
	\draw[shortout] (a3) -- (b3);
	\draw[shortout] (a4) -- (b4);

	\draw[thick,shorten <=-1mm] (u) -- (d) ;
	\draw[thick,shorten >=-1mm] (d) -- (v);
\end{tikzpicture}
\end{center}
The adjoint of the map $U_k^* : B_{k-1} \otimes V_k \to B_k$, pictured as the blue triangle on the right hand side, is then defined to be $U_k:B_k \to B_{k-1}\otimes V_k$ and becomes the next tensor in the MPS decomposition.   To check that the inductive hypothesis is satisfied, note that
$\alpha_{k} U_1 \cdots U_{k-1} U_k U_k^* = \alpha_{k-1} U_1 \cdots U_{k-1}$ since $\alpha_{k-1} U_1 \cdots U_{k-1} = W_k D_k U_k^*$ and $U_k^* U_k =1$.  Here is the picture proof:
\begin{center}
	\begin{tikzpicture}[x=1cm,y=1.5cm,baseline={(current bounding box.center)}]
        \node[hugetensor,fill=red!10, minimum width=90mm] (v15) at (4,1) {};
        \node[bigtensor, label = below left : {$V_1$}] (j0) at (0,2) {};
        \node[bigtensor] (j1) at (1,2) {};
        \node[] (j2) at (2,2) {$\cdots$};
        \node[bigtensor] (j3) at (3,2) {};
        \node[bigtensor,fill=blue!20, label = above right : $B_k$] (j4) at (4,2) {};
        \node[bigtensor,fill=blue!20,label = below right : $V_k$] (j5) at (5,2) {};
        \node[] (j6) at (6,1) {};
        \node[] (j7) at (7,1) {};
        \node[] (j8) at (8,1) {};

        \node[] (n0) at (0,1) {};
        \node[] (n1) at (1,1) {};
        \node[] (n2) at (2,1) {};
        \node[] (n3) at (3,1) {};
        \node[] (n4) at (4,1) {};
        \node[] (n5) at (5,1) {};
        \node[] (n6) at (6,0) {$V_{k+1}$};
        \node[] (n7) at (7,0) {$\cdots$};
        \node[] (n8) at (8,0) {$V_N$};

        \draw [shortin] (j0) -- (n0);
        \draw [shortin] (j1) -- (n1);
        \draw [shortin] (j3) -- (n3);
        \draw [shortin] (j4) -- (n4);
        \draw [shortout,shorten <=4mm] (n5) -- (j5);
        \draw [shortout] (j6) -- (n6);
        \draw [shortout] (j8) -- (n8);

        \draw[thick] (j0) -- (j1) -- (j2) -- (j3) -- (j4) -- (j5);
        \node[label = above left : {$B_{k-1}$}] (k6) at (6.5,2) {};
        \draw[witharrow] (j5) -- (k6);

        \node at (3.5,2.7) {$B_{k-1}$};
        \draw[->,thick2,opacity = .5] (3.5,2.5) -- (3.5,2.2) {};
\end{tikzpicture}
\end{center}
is equal to this
\begin{center}
\begin{tikzpicture}[x=1cm,y=1.5cm,baseline={(current bounding box.center)}]
	\node[medtriangle,minimum height = 8mm,fill = blue!20] (u3) at (0,4) {};
	\node[medtriangle,minimum height = 8mm,fill = blue!20,shape border rotate = 90] (u2) at (0,3) {};
	\node[medtriangle,minimum height = 8mm] (u) at (0,2) {};
	\node[smalltensor, minimum height = 2mm, fill = black!10] (d) at (0,1.25) {};
	\node[bigtriangle] (v) at (0,.5) {};

	\node (a1) at (-.25,2) {};
	\node (a2) at (.25,2) {};
	\node (a3) at (-1,.4) {};
	\node (a4) at (1,.4) {};
	\node[label = below left : $B_{k-1}$] (a5) at (-.25,5) {};
	\node[label = below right : $V_k$] (a6) at (.25,5) {};

	\node (b1) at (-.25,2.9) {};
	\node (b2) at (.25,2.9) {};
	\node (b3) at (-1,-.75) {$V_{k+1}$};
	\node (b4) at (1,-.75) {$V_N$};
	\node (b5) at (-.25,4) {};
	\node (b6) at (.25,4) {};

	\node at (0,-.25) {$\cdots$};

	\draw[shortin] (b1) -- (a1);
	\draw[shortin] (b2) -- (a2);
	\draw[shortout] (a3) -- (b3);
	\draw[shortout] (a4) -- (b4);
	\draw[shortin] (a5) -- (b5);
	\draw[shortin] (a6) -- (b6);

	\draw[thick,shorten <=-1mm,shorten >=-1mm] (u3) -- (u2);
	\draw[thick,shorten <=-1mm] (u) -- (d);
	 \draw[thick,shorten >= -1mm] (d) -- (v);

	\node at (3,2) {$=$};

	\node[medtriangle,minimum height = 8mm,fill = blue!20] (u) at (6,2) {};
	\node[smalltensor, minimum height = 2mm, fill = black!10] (d) at (6,1.25) {};
	\node[bigtriangle] (v) at (6,.5) {};

	\node (a1) at (5.75,2) {};
	\node (a2) at (6.25,2) {};
	\node (a3) at (5,.4) {};
	\node (a4) at (7,.4) {};
	\node (a5) at (6,5) {};

	\node[label = below left : $B_{k-1}$] (b1) at (5.75,2.9) {};
	\node[label = below right : $V_k$] (b2) at (6.25,2.9) {};
	\node (b3) at (5,-.75) {$V_{k+1}$};
	\node (b4) at (7,-.75) {$V_N$};
	\node (b5) at (6,4) {};

	\node at (6,-.25) {$\cdots$};

	\draw[shortin] (b1) -- (a1);
	\draw[shortin] (b2) -- (a2);
	\draw[shortout] (a3) -- (b3);
	\draw[shortout] (a4) -- (b4);

	\draw[thick,shorten <= -1mm] (u) -- (d);
	\draw[thick,shorten >= -1mm] (d) -- (v);
	\draw[rounded corners,thick2, dashed, gray] (-1, 1.5) rectangle (1, 3.5) {};
\end{tikzpicture}
\end{center}
which is the first picture:
\begin{center}
\begin{tikzpicture}[x=1cm,y=1.5cm,baseline={(current bounding box.center)}]
	\node[hugetensor,fill=red!10, minimum width=90mm] (v15) at (4,1) {};
	\node[bigtensor, label = below left : {$V_1$}] (j0) at (0,2) {};
	\node[bigtensor] (j1) at (1,2) {};
	\node[] (j2) at (2,2) {$\cdots$};
    \node[bigtensor] (j3) at (3,2) {};
    \node (j4) at (4.5,2) {};
    \node[] (j5) at (5.5,2) {};
    \node[] (j6) at (6,1) {};
    \node[] (j7) at (7,1) {};
    \node[] (j8) at (8,1) {};

    \node[] (n0) at (0,1) {};
    \node[] (n1) at (1,1) {};
    \node[] (n2) at (2,1) {};
    \node[] (n3) at (3,1) {};
    \node[] (n4) at (4,1) {};
    \node[] (m5) at (5,2) {$V_{k}$};
    \node[] (n5) at (5,1) {};
    \node[] (n6) at (6,0) {$V_{k+1}$};
    \node[] (n7) at (7,0) {$\cdots$};
    \node[] (n8) at (8,0) {$V_N$};

    \draw [shortin] (j0) -- (n0);
    \draw [shortin] (j1) -- (n1);
    \draw [shortin] (j3) -- (n3);
    \draw [shortin] (m5) -- (n5);
    \draw [shortout] (j6) -- (n6);
    \draw [shortout] (j8) -- (n8);

    \draw[thick] (j0) -- (j1) -- (j2) -- (j3);
    \node[label = above left : {$B_{k-1}$}] (k5) at (4.5,2) {};
    \draw[witharrow] (j4) -- (j3);
\end{tikzpicture}
\end{center}

In our application, the vector $|\psi\>$ and the operators $\beta_{k-1}: B_{k-1} \otimes V_k \to V_{k+1}\otimes \cdots \otimes V_N$ operate in spaces of such high dimensions that neither they, nor a direct SVD of them, is feasible.  Nonetheless, the $U_k$ operators can be obtained from an SVD of a reduced density operating in the effective space $B_{k-1}\otimes V_k$
\[\beta_{k-1}^* \beta_{k-1} : B_{k-1}\otimes V_k \to B_{k-1}\otimes V_k\]
In our application, the effective reduced density $\beta_{k-1}^* \beta_{k-1}$ can be computed as a double sum over the training examples and we can efficiently compute the tensors required for the inductive steps.  Then in the final step, the complementary space is small so the final map $U_N D_N:B_{N-1} \to V_N$ completes the reconstruction.  

More specifically, to define the $U_k$, we only need an eigenvector decomposition of $\beta_{k-1}^* \beta_{k-1}$, which looks like 
\begin{center}
\begin{tikzpicture}[x=1cm,y=1.5cm,baseline={(current bounding box.center)}]
	\node[hugetensor,fill=red!10, minimum width=90mm] (v15) at (4,1) {};
	\node[hugetensor,fill=red!10, minimum width=90mm] () at (4,-0.5) {};
	\node[bigtensor, label = below left : {$V_1$}] (j0) at (0,2) {};
	\node[bigtensor] (j1) at (1,2) {};
	\node[] (j2) at (2,2) {$\cdots$};
    \node[bigtensor] (j3) at (3,2) {};
    \node (j4) at (4.5,2) {};
    \node[] (j5) at (5.5,2) {};
    \node[] (j6) at (6,1) {};
    \node[] (j7) at (7,1) {};
    \node[] (j8) at (8,1) {};

    \node[] (n0) at (0,1) {};
    \node[] (n1) at (1,1) {};
    \node[] (n2) at (2,1) {};
    \node[] (n3) at (3,1) {};
    \node[] (n4) at (4,1) {};
    \node[] (m5) at (5,2) {$V_{k}$};
    \node[] (n5) at (5,1) {};
    \node[] (n6) at (6,0) {};
    \node[] (n7) at (7,0) {};
    \node[] (n8) at (8,0) {};

    \node[] (p5) at (5,-0.5) {};
    \node[] (p6) at (6,-0.5) {};
    \node[] (p7) at (7,-0.5) {};
    \node[] (p8) at (8,-0.5) {};
   	\node[] (p0) at (0,-0.5) {};
    \node[] (p1) at (1,-0.5) {};
    \node[] (p2) at (2,-0.5) {};
    \node[] (p3) at (3,-0.5) {};
    \node[] (p4) at (4,-0.5) {};

    \draw [shortin] (j0) -- (n0);
    \draw [shortin] (j1) -- (n1);
    \draw [shortin] (j3) -- (n3);
    \draw [shortin] (m5) -- (n5);
    \draw [thick,shorten <= 1.5mm,shorten >= 1.5mm] (p6) -- (j6);
    \draw [thick,shorten <= 1.5mm,shorten >= 1.5mm] (p8) -- (j8);
    
    \node[] (q5) at (5,-1.5) {$V_{k}$};
    \draw [shortout] (p5) -- (q5);

    \node[bigtensor, label = above left : {$V_1$}] (q0) at (0,-1.5) {};
	\node[bigtensor] (q1) at (1,-1.5) {};
	\node[] (q2) at (2,-1.5) {$\cdots$};
    \node[bigtensor] (q3) at (3,-1.5) {};
    \node (q4) at (4.5,-1.5) {};
    \node[label = below left : {$B_{k-1}$}] (q5) at (4.5,-1.5) {};
    \draw[witharrow] (q4) -- (q3);
	\draw[thick] (q0) -- (q1) -- (q2) -- (q3);	
    \draw[thick] (j0) -- (j1) -- (j2) -- (j3);
    \node[label = above left : {$B_{k-1}$}] (k5) at (4.5,2) {};
    \draw[witharrow] (j4) -- (j3);

    \draw [shortin] (q0) -- (p0);
    \draw [shortin] (q1) -- (p1);
    \draw [shortin] (q3) -- (p3);
    \node[] at (7,.25) {$\cdots$};
\end{tikzpicture}
\end{center}
and is given by a formula like the one in Equation \eqref{mainequation}.

In general, when factoring an arbitrary vector as an MPS, the bond spaces $B_{k}$ grow large exponentially fast.  Therefore, we may characterize data sets for which the MPS model is a good model by saying that $|\psi\>$ as defined in Equation \eqref{rho_pi} has an MPS model whose bond spaces $B_{k}$ remain small.  Alternatively, one can truncate or restrict the dimensions of the spaces $B_k$ resulting in a low rank MPS approximation of $|\psi\>$. As a criterion for this truncation, one can inspect the singular values at each inductive step and discard those which are small according to a pre-determined cutoff, and the corresponding columns of $U$ and $W$.  In the even-parity dataset that we investigate as an example, we always truncate $B_k$ to two dimensions throughout.

To understand whether this kind of low-rank approximation is useful, remember that we understand that the eigenvectors and eigenvalues of the reduced densities carry the essential prefix-suffix interactions.  
By having a training algorithm that emphasizes these eigenvalues and eigenvectors as the most important features of the data throughout training, the resulting model should be interpreted as capturing the most important prefix-suffix interactions.  We view these prefix-suffix interactions a proxy for the meaning of substrings within a language of larger strings.

\section{Under the hood}
With an in-depth understanding of the training algorithm, we aim to predict experimental results, simply given the fraction $0<f\leq 1$ of training samples used. Such an under-the-hood analysis shows that each tensor within the MPS is comprised of eigenvectors of a reduced density operator. The eigenvectors can be understood in terms of the reduced density matrix representation, which contains information from errors accrued in the algorithm's prior steps, along with combinatorial information from the current step. We now describe these ideas in careful detail. 


As an example, we perform an analysis of how well the algorithm learns on the even-parity dataset.  Let $\Sigma=\{0,1\}$ and consider the set $\Sigma^N$ of bitstrings of a fixed length $N$.  Define the \emph{parity} of a bitstring $(b_1, \ldots, b_N)$ to be
\begin{equation}
\parity(b_1, \ldots, b_N) := \sum_{i=1}^N b_i \mod 2.
\end{equation}
The set $\Sigma^N$ is partitioned into even and odd bitstrings:
\[E^N=\{s \in \Sigma^N:\parity(s)=0\} \text{ and }O^N=\{s\in \Sigma^N: \parity(s)=1\}\]
Consider the probability distribution $\pi:\Sigma^N\to \mathbb{R}$ uniformly concentrated on $E^N$:
\[ 
\pi(x) = \begin{cases} \frac{1}{2^{N-1}} & \text{ if $x\in E^N$} \\ 0 & \text{ if $x\in O^N$.}
\end{cases}
\]
This distribution defines a density $\rho_{\pi}=|E_N\>\<E_N|$ where 
\begin{equation}
\ket{E_N}=\frac{1}{\sqrt{2^{N-1}}}\sum_{s\in E^N}\ket{s}\in V_1\otimes V_2\otimes \cdots \otimes V_N
\end{equation} where $V_j\cong \mathbb{C}^2$ is the site space spanned by the bits in the $j$-th position. 
Choose a subset $T=\{s_1,\ldots, s_{N_T}\}\subset E_N$ of even parity bitstrings and let $f=N_T/2^{N-1}$ be the fraction selected.  The empirical distribution on this set defines the vector $|\psi\>= \frac{1}{\sqrt{N_T}} \sum_{i=1}^{N_T} |s_i\>$ as in Equation \eqref{emp_psi}. To begin our analysis on $|\psi\>$, let us closely inspect the algorithm's second step. The ideas therein will generalize to subsequent steps.

In step 2, we view each sample $s$ as a prefix-suffix pair $(a,b)$ where $a\in\Sigma^2$ and $b\in \Sigma^{N-2}$. We visualize the training set $T$ as a bipartite graph. Vertices represent prefixes $a$ and suffixes $b$ and there is an edge joining $a$ and $b$ if and only if $(a,b)\in T$.

\begin{minipage}{.5\textwidth}
\begin{center}
	\begin{tikzpicture}
	\node[vertex, label = left : {$00$}] (x1) at (0,.25) {};
	\node[vertex, label = left : {$11$}] (x2) at (0,-.25) {};

	\node[vertex, label = right : {$0000$}] (y1) at (1,1.5) {};
	\node[vertex, label = right : {$1100$}] (y2) at (1,1) {};
	\node[vertex, label = right : {$0110$}] (y3) at (1,.5) {};
	\node[vertex, label = right : {$0011$}] (y4) at (1,0) {};
	\node[vertex, label = right : {$1010$}] (y5) at (1,-.5) {};
	\node[vertex, label = right : {$0101$}] (y6) at (1,-1) {};
	\node[vertex, label = right : {$1001$}] (y7) at (1,-1.5) {};
	\node[vertex, label = right : {$1111$}] (y8) at (1,-2) {};

	\draw (x1) -- (y1) {};
	\draw (x1) -- (y2) {};
	\draw (x1) -- (y5) {};
	\draw (x1) -- (y6) {};
	\draw (x1) -- (y8) {};

	\draw (x2) -- (y1) {};
	\draw (x2) -- (y4) {};
	\draw (x2) -- (y6) {};
	\draw (x2) -- (y8) {};
	\end{tikzpicture}
\end{center}
\end{minipage}
\begin{minipage}{.5\textwidth}
\begin{center}
	\begin{tikzpicture}
	\node[vertex, label = left : {$01$}] (x1) at (0,.25) {};
	\node[vertex, label = left : {$10$}] (x2) at (0,-.25) {};

	\node[vertex, label = right : {$1000$}] (y1) at (1,1.5) {};
	\node[vertex, label = right : {$0100$}] (y2) at (1,1) {};
	\node[vertex, label = right : {$0010$}] (y3) at (1,.5) {};
	\node[vertex, label = right : {$0001$}] (y4) at (1,0) {};
	\node[vertex, label = right : {$0111$}] (y5) at (1,-.5) {};
	\node[vertex, label = right : {$1011$}] (y6) at (1,-1) {};
	\node[vertex, label = right : {$1101$}] (y7) at (1,-1.5) {};
	\node[vertex, label = right : {$1110$}] (y8) at (1,-2) {};
	
	\draw (x1) -- (y1);
	\draw (x1) -- (y3);
	\draw (x1) -- (y5);

	\draw (x2) -- (y3);
	\draw (x2) -- (y5);
	\draw (x2) -- (y6);
	\draw (x2) -- (y8);
	\end{tikzpicture}
\end{center}
\end{minipage}
Notice that samples in the left graph are concatenations of even parity bitstrings; samples in the right graph are concatenations of odd parity bitstrings. Let $\ket{\psi_2}\in \mathbb{C}^{\Sigma^2}\otimes \mathbb{C}^{\Sigma^{N-2}}$ denote the sum of the samples after having completed step 1,

\begin{equation}
	\begin{tikzpicture}[x=1cm,y=1.5cm,baseline={(current bounding box.center)}]
		\node[] at (-2,1) {$\ket{\psi_2}=$};
        \node[hugetensor,fill=red!10, minimum width=60mm,minimum height = 6mm] (v15) at (2.5,1) {};
        \node[bigtensor] (j0) at (0,1.75) {};
        \node[] (j1) at (1,2.5) {};
        \node[] (j2) at (2,2.5) {};
        \node[] (j3) at (3,2.5) {};
        \node[] (j4) at (4,2.5) {};
        \node[] (j5) at (5,2.5) {};

        \node[] (n0) at (0,1) {};
        \node[] (n1) at (1,1) {};
        \node[] (n2) at (2,1) {};
        \node[] (n3) at (3,1) {};
        \node[] (n4) at (4,1) {};
        \node[] (n5) at (5,1) {};

        \node[] (m0) at (0,2.5) {};

        \draw [shortin] (j0) -- (n0);
        \draw [shortin] (j1) -- (n1);
        \draw [shortin] (j2) -- (n2);
        \draw [shortin] (j3) -- (n3);
        \draw [shortin] (j4) -- (n4);
        \draw [shortin] (j5) -- (n5);

        \draw [witharrow] (m0) -- (j0);
\end{tikzpicture}
\end{equation}
and consider the reduced density $\rho_2=tr_{\Sigma^{N-2}}|\psi_2\>\<\psi_2|$. The entries of its matrix representation are understood from the data in the graph. Choosing an ordering on the set $\Sigma^2$, we write $\rho_2$ as

\begin{equation}\label{eq:rho2}
\rho_2\;=\;
\frac{1}{N_T}
\begin{bmatrix}
d_1 & s_e & 0 & 0 \\
s_e & d_2 & 0 & 0 \\
0 & 0 & d_3 & s_o \\
0 & 0 & s_o & d_4
\end{bmatrix}
\end{equation}
The number of training samples $N_T$ is the total number of edges in the graph. The diagonal entries are the degrees of vertices associated to prefixes: $d_1$ is the degree of 00, $d_2$ is the degree of 11, $d_3$ is the degree of 01, $d_4$ is the degree of  10.  The off-diagonal entries are the number of paths of length 2 in each component of the graph. That is, $s_e$ is the number of suffixes that $00$ and $11$ have in common; $s_o$ is the number of suffixes that $01$ and $10$ have in common.  If $T$ contains all samples then both graphs are complete bipartite and the entries of $\rho_2$ are all equal (to $2^{N-3}$ in this case).  In this case, $\rho_2$ is a rank 2 operator.  It has two eigenvectors---one from each block.  This is the idealized scenario: every sequence is present in the training set, the tensor obtained $\rho_2=U_2D_2U_2^*$ is
then 
\[\rho_2=\frac{1}{2}(|E_2\>\<E_2| \oplus |O_2\>\<O_2|)\]
where $\ket{E_{2}}=\frac{1}{\sqrt{2}}(|00\> + |11\>)$ denotes the normalized sum of even prefixes of length $2$, and $\ket{O_{2}}=\frac{1}{\sqrt{2}}(|01\> + |10\>)$ denotes the normalized sum of odd prefixes of length $2$. 
As a matrix, $U_2$ has $\ket{E_2}$ and $\ket{O_2}$ along its rows.
We think of it as a ``summarizer'':  
it projects a prefix onto an axis that can be identified with either $|E_2\>$ or $|O_2\>$ according to its parity, perfectly summarizing the information of that prefix required to understand which suffixes it is paired with.

More generally, however, if $T \neq E_N$ then the reduced density $\rho_2$ may be full rank. In this case we choose the eigenvectors $\ket{E'_2}, \ket{O'_2}$ that correspond to the two largest eigenvalues of $\rho_2$. We assume these eigenvectors come from distinct blocks. This defines the tensor $U_2$, which as a matrix has $|E'_2\>$ and $|O'_2\>$ along its rows, where
\begin{align*}
\ket{E'_2}&=\cos\theta_2\ket{00} + \sin\theta_2\ket{11} \\
\ket{O'_2}&=\cos\phi_2\ket{01}  + \sin\phi_2\ket{10} 
\end{align*}
for some angles $\theta_2$ and $\phi_2$. These angles can be computed following the expression in \eqref{evects} for the eigenvectors: 
\[
\theta_2 = \arctan\left(\frac{2s_e}{\sqrt{G_e^2 + 4s_e^2} + G_e}\right) \quad\text{and}\quad
\phi_2= \arctan\left(\frac{2s_o}{\sqrt{G_o^2 + 4s_o^2} + G_o}\right)
\]
Here, $G_e=d_1-d_2$ and $G_o=d_3-d_4$  denote the gaps between the diagonal entries in each block. The angles should be thought of as measuring the deviation from perfect learning in step 2: if $f=1$ then $G_e,G_o=0$ and so $\theta_2=\phi_2=\pi/4$ which implies $\ket{E'_2}=\ket{E_2}$ and $\ket{O'_2}=\ket{O_2}$. In this case, step 2 has worked perfectly.   Note that this is not an if-and-only-if scenario.  Even if $f<1$ then the reduced density may \emph{still} have $\ket{E_2}$ and $\ket{O_2}$ as its eigenvectors. Indeed, this occurs whenever $G_e=G_o=0$ and $s_e,s_o\neq 0$.  In that case, the eigenvectors of $\rho_2$ are the desired parity vectors $\ket{E_2},\ket{O_2}$, and the summarizer $U_2$ obtained is a true summarization tensor. But if $G_e$ or $G_o$ are both nonzero, then step 2 induces a summarization error, which we measure as the deviation of $\theta_2$ and $\phi_2$ from the desired $\pi/4$. 

The analysis described here is repeated at each subsequent step $k=3,\ldots,N$, with minor adjustments to the combinatorics. So let us now describe the general schema. In the $k$th step of the training algorithm, each sample is cut after the $k$-th bit and viewed as a prefix-suffix pair $s=(a,b)$ where $a\in \Sigma^k$ and $b\in \Sigma^{N-k}$. Let $|\psi_k\>\in \mathbb{C}^{\Sigma^k}\otimes \mathbb{C}^{\Sigma^{N-k}}$ denote the sum of the samples after having completed step $k-1$.

\begin{center}
	\begin{tikzpicture}[x=1cm,y=1.5cm,baseline={(current bounding box.center)}]
		\node[] at (-2,1.75) {$\ket{\psi_3}=$};
        \node[hugetensor,fill=red!10, minimum width=60mm,minimum height = 6mm] (v15) at (2.5,1) {};
        \node[bigtensor] (j0) at (0,1.75) {};
        \node[bigtensor] (j1) at (1,1.75) {};
        \node[] (j2) at (2,2.5) {};
        \node[] (j3) at (3,2.5) {};
        \node[] (j4) at (4,2.5) {};
        \node[] (j5) at (5,2.5) {};

        \node[] (n0) at (0,1) {};
        \node[] (n1) at (1,1) {};
        \node[] (n2) at (2,1) {};
        \node[] (n3) at (3,1) {};
        \node[] (n4) at (4,1) {};
        \node[] (n5) at (5,1) {};

        \node[] (m1) at (1,2.5) {};

        \draw [shortin] (j0) -- (n0);
        \draw [shortin] (j1) -- (n1);
        \draw [shortin] (j2) -- (n2);
        \draw [shortin] (j3) -- (n3);
        \draw [shortin] (j4) -- (n4);
        \draw [shortin] (j5) -- (n5);

        \draw[witharrow] (m1) -- (j1);
        \draw[thick] (j0) -- (j1);
\end{tikzpicture}
\end{center}
and let $\rho_k:=tr_{\Sigma^{N-k}}|\psi_k\>\<\psi_k|$ denote the reduced density on the prefix subsystem at step $k$. It is an operator on $B_{k-1}\otimes V_k$, where $B_{k-1}$ is a 2-dimensional space which may be identified with the span of the eigenvectors associated to the two largest eigenvalues of $\rho_{k-1}.$ As a matrix, $\rho_k$ is a direct sum of $2\times 2$ matrices,
\begin{equation}\label{eq:rhok}
\rho_k\;=\;
\frac{1}{tr(\rho_k)}
\begin{bmatrix}
e0 & s_e & 0 & 0\\
s_e & o1 & 0 & 0\\
0 & 0 & e1 & s_o\\
0 & 0 & s_o & o0
\end{bmatrix}
\end{equation}
We postpone a description of the entries until Section \ref{sec:comb}. But know that, as in the case when $k=2$, the upper and lower blocks contains combinatorial information about prefixes of even and odd parity, respectively. As before, we are interested in the largest eigenvectors $\ket{E_k'},\ket{O_k'}$ contributed by each block. They define the tensor $U_k$, which as a matrix has $|E'_k\>$ and $|O'_k\>$ along its rows, and can be understood inductively. The eigenvectors contain combinatorial information from step $k$ along with data from step $k-1$. Let $\ket{E'_1}:=\ket 0$ and $\ket{O'_1} := \ket 1$. Then for $k\geq 2$
 
\begin{align*}
\ket{E'_k}&=\cos\theta_k\ket{E'_{k-1}}\otimes \ket{0} + \sin\theta_k\ket{O'_{k-1}}\otimes \ket 1\\
\ket{O'_k}&=\cos\phi_k\ket{E'_{k-1}}\otimes \ket{1} + \sin\phi_k\ket{O'_{k-1}}\otimes \ket 0
\end{align*}
where 
\begin{equation}\label{eq:angles}
\theta_k = \arctan\left(\frac{2s_e}{\sqrt{G_e^2 + 4s_e^2}+G_e}\right)
\quad
\phi_k = \arctan\left(\frac{2s_o}{\sqrt{G_o^2 + 4s_o^2}+G_o}\right)
\end{equation}
Again, the angles are a measurement of the error accrued in step $k$.  Significantly, no error is accrued when the gaps $G_e:=e0-o1$ and $G_o:=e1-o0$ are zero and the off-diagonals $s_e,s_o$ are non-zero, for then $\theta_k=\phi_k=\pi/4$.  This outcome, or one close to it, is statistically favored for a wide range of training fractions.

As a matrix, 
\[
U_k =\begin{bmatrix}
\cos\theta_k & \sin\theta_k & 0 & 0\\
0 & 0 & \cos\phi_k & \sin\phi_k
\end{bmatrix}
\]
and so $U_k$ is akin to a map $B_{k-1}\otimes V_k\to B_k$ that combines previously summarized information from $B_{k-1}$ with new information from $V_k$. It then summarizes the resulting data by projecting onto one of two orthogonal vectors, which may be identified with $\ket{E'_k}$ or $\ket{O'_k}$, in the new bond space $B_k$.
\begin{center}
	\begin{tikzpicture}
    \node[bigtensor, label = above : {$U_3$}] (u) at (4.5,4) {};
    \node[] (u0) at (3,4) {$|01\>$};
    \node[] (u2) at (6,4) {$|E'_3\>$};
    \node[] (n1) at (4.5,2.5) {$|1\>$};
    \draw[witharrow] (n1)--(u);
    \draw[witharrow]  (u)--(u2);
    \draw[witharrow] (u0) -- (u);
 \end{tikzpicture}
\end{center}
The true orientation of the arrows on $U_k$ are down-left, rather than up-right. But the vector spaces in question are finite-dimensional, and our standard bases provide an isomorphism between a space and its dual. That is, no information is lost by momentarily adjusting the arrows for the purposes of sharing intuition.

In summary, this template provides a concrete handle on the tensors $U_k$ that comprise the MPS factorization of $|\psi\>$.

\subsection{High-level summary}
We close by summarizing the high-level ideas present in this under-the-hood analysis.  At the $k$th step of the training algorithm one obtains a $4\times 4$ block diagonal reduced density matrix $\rho_k$. It is given in Equation \eqref{eq:rho2} in the case when $k=2$ and as in Equation \eqref{eq:rhok} when $k>2$. These matrices are obtained by tracing out the suffix subsystem from the projection $|\psi_k\>\<\psi_k|$, where $|\psi_k\>$ is the sum of the samples in the training set after having completed step $k-1$. Since $|\psi_k\>$ depends on the error obtained in step $k-1$, so does $\rho_k$. This error is defined by the angles $\theta_{k-1}$ and $\phi_{k-1}$. As shown in Equation \eqref{eq:angles}, these angles---and hence the error---are functions of the entries of the matrix representing $\rho_{k-1}$.  
So, the $k$th level density takes into account the errors accrued at each subsequent step as well as combinatorial information in the present step.  A partial trace computation thus directly leads to the matrix representation for $\rho_k$ given in Equation \eqref{eq:rhok}.
Explicitly, the non-zero entries of the matrix are computed by Equations \eqref{eq:e0} and \eqref{eq:se}. With this, one has full knowledge of the matrix $\rho_k$ and therefore of its eigenvectors $|E_k'\>,|O_k'\>$. Written in the computational basis, they are of the form shown in Equation \eqref{evects}. These two eigenvectors then assemble to form the rows of the tensor $U_k$, when viewed as a $2\times 4$ matrix. 

This analysis gives a thorough understanding of the error propagated at each step of the algorithm, as well as of the final MPS $|\psi_{\text{MPS}}\>$. To measure the algorithm's performance, we begin by evaluating the inner product of this vector with an MPS decomposition of the target vector $|E_N\>$.

\begin{center}
	\begin{tikzpicture}
	\node[] at (-2,.5) {$\<E_N|\psi_{\text{MPS}}\> = $};
	\node[bigtensor] (p0) at (0,1.5) {};
	\node[bigtensor] (p1) at (1,1.5) {};
	\node[bigtensor] (p2) at (2,1.5) {};
	\node[bigtensor] (p3) at (3,1.5) {};
	\node[bigtensor] (p4) at (4,1.5) {};
	\node[bigtensor] (p5) at (5,1.5) {};

	\node[bigtensor,fill = green!20] (e0) at (0,0) {};
	\node[bigtensor,fill = green!20] (e1) at (1,0) {};
	\node[bigtensor,fill = green!20] (e2) at (2,0) {};
	\node[bigtensor,fill = green!20] (e3) at (3,0) {};
	\node[bigtensor,fill = green!20] (e4) at (4,0) {};
	\node[bigtensor,fill = green!20] (e5) at (5,0) {};

	\draw[thick] (p0) -- (p1) -- (p2) -- (p3) -- (p4) -- (p5);
	\draw[thick] (e0) -- (e1) -- (e2) -- (e3) -- (e4) -- (e5);

	\draw[witharrow] (p0) -- (e0);
	\draw[witharrow] (p1) -- (e1);
	\draw[witharrow] (p2) -- (e2);
	\draw[witharrow] (p3) -- (e3);
	\draw[witharrow] (p4) -- (e4);
	\draw[witharrow] (p5) -- (e5);

	\end{tikzpicture}
\end{center}

The $k$th tensor comprising the decomposition of $|E_N\>$ is equal to $U_k$ when $\theta_k$ and $\phi_k$ are evaluated at $\pi/4.$ The contraction thus results in a sum of products of $\cos\theta_k,\sin\theta_k,\cos\phi_k,\sin\phi_k$ for $k=2,\ldots,N$. More concretely, for each even bitstring $s\in E^N$ the inner product $\<s|\psi_{\text{MPS}}\>$ is the square root of the probability of $s$. For now, we'll refer to it as the \emph{weight} $w(s):=\<s|\psi_{\text{MPS}}\>$ associated to the sample $s$. For each $s$, its weight $w(s)$ is a product of various $\cos\theta_k,\sin\theta_k,\cos\phi_k,\sin\phi_k$, the details of which are given in Section \ref{sec:comb}. The final overlap is then the sum  

\begin{equation}\label{eq:estimate}
\<E_N|\psi_{\text{MPS}}\> = \frac{1}{\sqrt{2^{N-1}}}\sum_{s\in E^N}w(s)
\end{equation}

\begin{center}
	\begin{figure}[h]
	\begin{subfigure}[t]{0.7\textwidth}
	\includegraphics[scale=0.6]{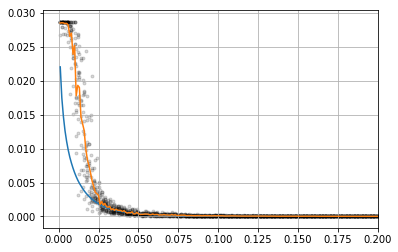}
	\caption{The experimental average (orange) and theoretical prediction (blue).}
	\end{subfigure} 
	\begin{subfigure}[t]{0.7\textwidth}
	\includegraphics[scale=0.6]{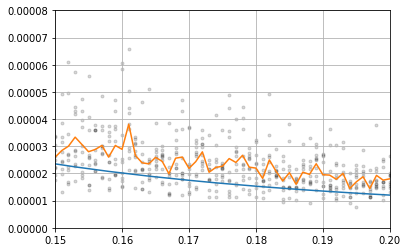}
	\caption{A closer look for $0.15\leq f \leq 0.2.$}
	\end{subfigure}
	\caption{The experimental average (orange) and theoretical prediction (blue) of the weighted Bhattacharya distance between the probability distribution learned experimentally and the theoretical prediction for bit strings of length $N=16$ and training set fractions of $0<f\leq 0.2$.}\label{fig:experiments}
	\end{figure}
	\end{center}

Now, suppose the training set consists of a fraction $f$ of the entire population.  The entries of the reduced densities in \eqref{eq:rhok} are described combinatorially, as detailed in the next section.  This makes it possible to make statistical estimates for gaps $G_e$ and $G_o$ and off-diagonal entries $s_o$ and $s_e$ in \eqref{eq:angles}.  Therefore, we can make statistical predictions for the angles $\theta_k$ and $\phi_k$ and hence for the tensors $U_k$ comprising the trained MPS and the resulting generalization error.  The results are plotted in Figure \ref{fig:experiments}, where we use the Bhattacharya distance
\begin{equation}\label{eq:bhatt}
    -\frac{1}{\sqrt{2^{N-1}}}\ln\left(\sum_{s\in E^N}w(s)\right)
\end{equation}
between the true population distribution and the one defined by either an experimentally trained MPS as a proxy for generalization error. The theoretical curve could, in principle, be improved by making more accurate statistical estimates for the combinatorics involved.  

\subsection{Combinatorics of reduced densities}\label{sec:comb}
We now describe the entries of $k$th level reduced density in Equation \eqref{eq:rhok}. They depend on certain combinatorics in step $k$ as well as error accumulated in the previous step. The latter has an inductive description. To start, observe that the parity of a prefix $a\in\Sigma^k$ is determined by its last bit, together with the parity of its first $k-1$ bits. The set $\Sigma^k$ thus partitions into four sets:
\begin{align*}
E0 = \{a\in\Sigma^k:a=(e_{k-1},0) \text{ where $e_{k-1}\in E^{k-1}$}\}\\
O1 = \{a\in\Sigma^k:a=(o_{k-1},1) \text{ where $o_{k-1}\in O^{k-1}$}\}\\
E1 = \{a\in\Sigma^k:a=(e_{k-1},1) \text{ where $e_{k-1}\in E^{k-1}$}\}\\
O0 = \{a\in\Sigma^k:a=(o_{k-1},0) \text{ where $o_{k-1}\in O^{k-1}$}\}\\
\end{align*}
By viewing the training set as a bipartite graph, one has a visual understanding of these sets: $E0$ contains all prefixes of even parity whose last bit is 0; $O1$ contains all prefixes of even parity whose last bit is $1$, and so on. In the example below with $k=3$, we use color to distinguish each set.

\begin{minipage}{.5\textwidth}
\begin{center}
	\begin{tikzpicture}
	\node[vertex, label = left : {$00\;0$}] (x1) at (0,1.5) {};
	\node[vertex, label = left : {$11\;0$}] (x2) at (0,1){};
	\node[vertex, label = left : {$01\;1$}] (x3) at (0,.5){};
	\node[vertex, label = left : {$10\;1$}] (x4) at (0,0){};

	\node[vertex, label = right : {$000$}] (y1) at (1,1.5){};
	\node[vertex, label = right : {$110$}] (y2) at (1,1){};
	\node[vertex, label = right : {$101$}] (y3) at (1,.5){};
	\node[vertex, label = right : {$011$}] (y4) at (1,0){};

	\draw[OrangeRed,thick2] (x1) -- (y1) {};
	\draw[OrangeRed,thick2] (x1) -- (y3) {};
	\draw[OrangeRed,thick2] (x2) -- (y1) {};
	\draw[OrangeRed,thick2] (x2) -- (y3) {};
	\draw[OrangeRed,thick2] (x2) -- (y4) {};
	\draw[ProcessBlue,thick2] (x3) -- (y1) {};
	\draw[ProcessBlue,thick2] (x4) -- (y2) {};
	\draw[ProcessBlue,thick2] (x4) -- (y4) {};

	\node at (.5,1.75) {{\color{OrangeRed}$E0$}};
	\node at (.5,-.25) {{\color{ProcessBlue}$O1$}};

	\node at (-1.5,1.5) {${\color{gray}{\cos\theta_2}}$};
	\node at (-1.5,1) {${\color{gray}{\sin\theta_2}}$};
	\node at (-1.5,.5) {${\color{gray}{\cos\phi_2}}$};
	\node at (-1.5,0) {${\color{gray}{\sin\phi_2}}$};
	\end{tikzpicture}
\end{center}
\end{minipage}
\begin{minipage}{.5\textwidth}
\begin{center}
	\begin{tikzpicture}
	\node[vertex, label = left : {$00\;1$}] (x1) at (0,1.5) {};
	\node[vertex, label = left : {$11\;1$}] (x2) at (0,1){};
	\node[vertex, label = left : {$01\;0$}] (x3) at (0,.5){};
	\node[vertex, label = left : {$10\;0$}] (x4) at (0,0){};

	\node[vertex, label = right : {$100$}] (y1) at (1,1.5){};
	\node[vertex, label = right : {$010$}] (y2) at (1,1){};
	\node[vertex, label = right : {$001$}] (y3) at (1,.5){};
	\node[vertex, label = right : {$111$}] (y4) at (1,0){};

	\draw[ForestGreen,thick2] (x1) -- (y1);
	\draw[ForestGreen,thick2] (x1) -- (y2);
	\draw[ForestGreen,thick2] (x1) -- (y4);
	\draw[ForestGreen,thick2] (x2) -- (y4);
	\draw[YellowOrange,thick2] (x3) -- (y2);
	\draw[YellowOrange,thick2] (x3) -- (y4);
	\draw[YellowOrange,thick2] (x4) -- (y2);
	\draw[YellowOrange,thick2] (x4) -- (y4);

	\node at (.5,1.75) {{\color{ForestGreen}$E1$}};
	\node at (.5,-.25) {{\color{YellowOrange}$O0$}};

	\node at (-1.5,1.5) {${\color{gray}{\cos\theta_2}}$};
	\node at (-1.5,1) {${\color{gray}{\sin\theta_2}}$};
	\node at (-1.5,.5) {${\color{gray}{\cos\phi_2}}$};
	\node at (-1.5,0) {${\color{gray}{\sin\phi_2}}$};
	\end{tikzpicture}
\end{center}
\end{minipage}
As shown, each prefix also has a weight that records its contribution to the error accumulated in previous steps. Concretely, we assign to each prefix $a\in \Sigma^k$ a weight $w(a)$, which is a product of $k-2$ terms. For $2\leq i \leq k-1,$ the $i$th factor of $w(a)$ is defined to be
	\begin{itemize}
		\item $\cos\theta_i$ if the parity of the first $i-1$ bits is even and the $i$th bit is 0
		\item $\sin\theta_i$ if the parity of the first $i-1$ bits is odd and the $i$th bit is 1
		\item $\cos\phi_i$ if the parity of the first $i-1$ bits is even and the $i$th bit is 1
		\item $\sin\phi_i$ if the parity of the first $i-1$ bits is odd and the $i$th bit is 0
		\end{itemize}
For example, if $k=3$ then $w(011)=\cos\phi_2$. If $k=5$ then $w(01101)=\cos\theta_4\sin\theta_3\cos\phi_2$. These weights are naturally associated to each tensor. For instance, recalling that each tensor $U_k$ is akin to a summarizer, one sees $w(01101)$ in the following way:

\begin{center}
	\begin{tikzpicture}
    \node[bigtensor, label = above : {$U_2$}] (a) at (0,1) {};
    \node[] (aL) at (-1.5,1) {$\ket{0}$};
    \node[] (aR) at (2,1) {$\cos\phi_2\ket{01}$};
    \node[] (aB) at (0,-.5) {$\ket{1}$};

    \draw[witharrow] (aB)--(a);
    \draw[witharrow]  (a)--(aR);
    \draw[witharrow] (aL) -- (a);

    \node[bigtensor, label = above : {$U_3$}] (b) at (4,1) {};
    \node[] (bL) at (2.7,1) {};
    \node[] (bR) at (6.5,1) {$\sin\theta_3\cos\phi_2\ket{011}$};
    \node[] (bB) at (4,-.5) {$\ket{1}$};

    \draw[witharrow] (bB)--(b);
    \draw[witharrow]  (b)--(bR);
    \draw[witharrow] (bL) -- (b);

    \node[bigtensor, label = above : {$U_4$}] (c) at (9,1) {};
    \node[] (cL) at (7.7,1) {};
    \node[] (cR) at (12.2,1) {$\cos\theta_4\sin\theta_3\cos\phi_2\ket{0110}$};
    \node[] (cB) at (9,-.5) {$\ket{0}$};

    \draw[witharrow] (cB)--(c);
    \draw[witharrow]  (c)--(cR);
    \draw[witharrow] (cL) -- (c);

    \end{tikzpicture}
\end{center}



We can now describe the entries of the reduced density defined in Equation \eqref{eq:rhok}. The first diagonal entry is
\begin{equation}\label{eq:e0}
e0 = \sum_{\text{suffixes }b}\left(\sum_{\substack{a\in E0 \\ (a,b)\in T}} w(a)\right)^2
\end{equation}
and the other diagonals are defined similarly. If perfect learning occurs then $e0$ is, up to a normalizing constant, the sum of the squares of the degrees of each suffix, with respect to $E0$. For example, in the graph below $e0$ is proportional to $ 2^2 + 2^2 + 1^2=9$.

\begin{center}
\begin{minipage}{.5\textwidth}
	\begin{tikzpicture}

	\node[vertex, label = left : {$00\;0$}] (x1) at (0,1.5) {};
	\node[vertex, label = left : {$11\;0$}] (x2) at (0,1){};
	\node[vertex, label = left : {$01\;1$}] (x3) at (0,.5){};
	\node[vertex, label = left : $10\;1$] (x4) at (0,0){};

	\node[vertex, label = right : {$000$}] (y1) at (1,1.5){};
	\node[vertex, label = right : {$110$}] (y2) at (1,1){};
	\node[vertex, label = right : {$101$}] (y3) at (1,.5){};
	\node[vertex, label = right : {$011$}] (y4) at (1,0){};

	\draw[OrangeRed,thick2] (x1) -- (y1) {};
	\draw[OrangeRed,thick2] (x1) -- (y3) {};
	\draw[OrangeRed,thick2] (x2) -- (y1) {};
	\draw[OrangeRed,thick2] (x2) -- (y3) {};
	\draw[OrangeRed,thick2] (x2) -- (y4) {};
	\draw[ProcessBlue,opacity=.25] (x3) -- (y1) {};
	\draw[ProcessBlue,opacity=.25] (x4) -- (y2) {};
	\draw[ProcessBlue,opacity=.25] (x4) -- (y4) {};

	\node at (.5,1.75) {{\color{OrangeRed}$E0$}};

	\node at (-1.5,1.5) {${\color{gray}{\cos\theta_2}}$};
	\node at (-1.5,1) {${\color{gray}{\sin\theta_2}}$};
	\node at (-1.5,.5) {${\color{gray}{\cos\phi_2}}$};
	\node at (-1.5,0) {${\color{gray}{\sin\phi_2}}$};
	\end{tikzpicture}
\end{minipage}
\end{center}
In general, though, the summands will not be integers but rather products of weights.
The off-diagonal entry in the even block of the reduced density is 
\begin{equation}\label{eq:se}
s_e=\sum_{\text{suffixes }b}\left(\sum_{\substack{a\in E0, \;a'\in O1 \\ (a,b),(a',b)\in T}} w(a)\cdot w(a')\right)
\end{equation}
When perfect learning occurs, $s_e$ counts the number of paths of length 2, where now a path is comprised of one edge from $E0$ and one edge from $O1$. For example, in the graph below $s_e=3.$ 

\begin{center}
	\begin{tikzpicture}
	\node[vertex] (x1) at (0,1.5) {};
	\node[vertex] (x2) at (0,1){};
	\node[vertex] (x3) at (0,.5){};
	\node[vertex] (x4) at (0,0){};

	\node[vertex] (y1) at (1,1.5){};
	\node[vertex] (y2) at (1,1){};
	\node[vertex] (y3) at (1,.5){};
	\node[vertex] (y4) at (1,0){};

	\draw[OrangeRed,thick2] (x1) -- (y1);
	\draw[OrangeRed,thick2] (x2) -- (y1);
	\draw[OrangeRed,thick2] (x2) -- (y4);
	\draw[OrangeRed,opacity=.25] (x1) -- (y3);
	\draw[OrangeRed,opacity=.25] (x2) -- (y3);
	\draw[ProcessBlue,thick2] (x3) -- (y1);
	\draw[ProcessBlue,thick2] (x4) -- (y4);
	\draw[ProcessBlue,opacity=.25] (x4) -- (y2);

	\node at (1.5,.75) {$=$};

	\node[vertex] (a1) at (2,1.5) {};
	\node[vertex] (a2) at (2,1){};
	\node[vertex] (a3) at (2,.5){};
	\node[vertex] (a4) at (2,0){};

	\node[vertex] (b1) at (3,1.5){};
	\node[vertex] (b2) at (3,1){};
	\node[vertex] (b3) at (3,.5){};
	\node[vertex] (b4) at (3,0){};

	\draw[OrangeRed,opacity=.25] (a1) -- (b1);
	\draw[OrangeRed,opacity=.25] (a1) -- (b3);
	\draw[OrangeRed,opacity=.25] (a2) -- (b1);
	\draw[OrangeRed,opacity=.25] (a2) -- (b3);
	\draw[OrangeRed,thick2] (a2) -- (b4);
	\draw[ProcessBlue,opacity=.25] (a3) -- (b1);
	\draw[ProcessBlue,opacity=.25] (a4) -- (b2);
	\draw[ProcessBlue,thick2] (a4) -- (b4);

	\node at (3.5,.75) {$+$}; 

	\node[vertex] (c1) at (4,1.5) {};
	\node[vertex] (c2) at (4,1){};
	\node[vertex] (c3) at (4,.5){};
	\node[vertex] (c4) at (4,0){};

	\node[vertex] (d1) at (5,1.5){};
	\node[vertex] (d2) at (5,1){};
	\node[vertex] (d3) at (5,.5){};
	\node[vertex] (d4) at (5,0){};

	\draw[OrangeRed,thick2] (c1) -- (d1);
	\draw[OrangeRed,opacity=.25] (c2) -- (d1);
	\draw[OrangeRed,opacity=.25] (c2) -- (d3);
	\draw[OrangeRed,opacity=.25] (c2) -- (d4);
	\draw[ProcessBlue,thick2] (c3) -- (d1);
	\draw[ProcessBlue,opacity=.25] (c4) -- (d2);
	\draw[ProcessBlue,opacity=.25] (c4) -- (d4);

	\node at (5.5,.75) {$+$};

	\node[vertex] (u1) at (6,1.5) {};
	\node[vertex] (u2) at (6,1){};
	\node[vertex] (u3) at (6,.5){};
	\node[vertex] (u4) at (6,0){};

	\node[vertex] (v1) at (7,1.5){};
	\node[vertex] (v2) at (7,1){};
	\node[vertex] (v3) at (7,.5){};
	\node[vertex] (v4) at (7,0){};

	\draw[OrangeRed,opacity=.25] (u1) -- (v1);
	\draw[OrangeRed,thick2] (u2) -- (v1);
	\draw[OrangeRed,opacity=.25] (u2) -- (v3);
	\draw[OrangeRed,opacity=.25] (u2) -- (v4);
	\draw[ProcessBlue,thick2] (u3) -- (v1);
	\draw[ProcessBlue,opacity=.25] (u4) -- (v2);
	\draw[ProcessBlue,opacity=.25] (u4) -- (v4);

	\node at (.5,1.75) {{\color{OrangeRed}$E0$}};
	\node at (.5,-.25) {{\color{ProcessBlue}$O1$}};

	\end{tikzpicture}
\end{center}
In general, however, $s_e$ will be a sum of products of weights. The expression for the off-diagonal $s_o$ in the odd block is similar to that in Equation \eqref{eq:se}.

In summary, the theory behind the reduced densities and their eigenvectors gives us an exact understanding of the error propagated through each step of the training algorithm. We may then predict the Bhattacharya distance in \eqref{eq:bhatt} using statistical estimates of the expected combinatorics. This provides an accurate prediction based solely on the fraction $f$ of training samples used and the length $N$ of the sequences.

\section{Experiments}

The training algorithm was written in the ITensor library \cite{iTensor}; the code is available on Github. For a fixed fraction $0<f\leq0.2$ we run the algorithm on ten different datasets, each containing $N_T=f2^{N-1}$ bitstrings of length $N=16$. We then compare the average Bhattacharya distance in Equation \eqref{eq:bhatt} to the theoretical prediction. To handle the angles $\theta_k$ and $\phi_k$ in the theoretical model, we make a few simplifying assumptions about the expected behavior of the combinatorics. 

First we assume $\theta = \phi_k$ for all $k$ since the combinatorics of both blocks of the reduced densities $\rho_k$ in (\ref{eq:rhok}) have similar behavior. We further assume the average angle $\theta$ is a function of the average off-diagonal $s_e$ and the average diagonal gap $G_e$ at the $k$th step, that is $\mathbb{E}[\theta_k(s_e,G_e)] = \theta_k(\mathbb{E}[s_e],\mathbb{E}[G_e])$ for all $k$. The expectation for $s_e$ is experimentally determined to be independent of $k$, and dependent on the fraction $f$ and bitstring length $N$ alone: $\mathbb{E}[s_e] = f\cdot N_T/4$ for all $k$. We approximate the expected gap $G_e$ at the $k$th step to be an experimentally determined function of $f$ and the expected gap $G_2=|d_1-d_2|$ of the diagonal entries of the reduced density defined at step 2 of the algorithm.  Understanding the expected behavior of $G_2$ is similar to understanding the statistics of a coin toss. On average, one expects to flip the same number of heads and tails and yet the expectation for their difference is non-zero. The distribution for $G_2$ is similar, but a little different:
\begin{align*}
\mathbb{E}[G_2]=\sum_{d_1}|2d_1-r|\frac{\binom{n}{d_1}\binom{n}{r-d_1}}{\binom{2n}{r}}
\end{align*}
where $r=d_1+d_2=N_T/2$ and $n=2^{N-3}$ is the number of even parity bitstrings of length $N-2$. The plots in Figure \ref{fig:experiments} compare the theoretical estimate against the experimental average.

\section{Conclusion}


Models based on tensor networks open interesting directions for machine learning research. Tensor networks can be viewed as a sequence of related linear maps, which by acting together on a very high-dimensional space allows the model to be arbitrarily expressive. The underlying linearity and powerful techniques from linear algebra allow us to pursue a training algorithm where we can look ``under the hood'' to understand each step and its consequences for the ability of our model to reconstruct a particular data set, the even-parity data set.

Our work also highlights the advantages of working in a probability formalism based on the 2-norm. This is the same formalism used to interpret the wavefunction in quantum mechanics; here we use it as a framework to treat classical data. Density matrices naturally arise as the 2-norm analogue of marginal probability distributions familiar from conventional 1-norm probability. Marginals still appear as the diagonal of the density matrix. Unlike marginals, the density matrices we use hold sufficient information to reconstruct the entire joint distribution. Our training algorithm can be summarized as estimating the density matrix from the training data, then reconstructing the joint distribution step-by-step from these density matrix estimates.

The theoretical predictions we obtained for the generative performance of the model agree well with the experimental results. Note that care is needed to compare these results, since the theoretical approach involves averaging over all possible training sets to produce a single typical weight MPS, whereas the experiments produce a different weight MPS for each training-set sample. In the near future, we look forward to extending our approach to other measures of model performance and behavior, and certainly other data sets as well. 

More ambitiously, we hope this work points the way to theoretically sound and robust predictions of machine learning model performance based on empirical summaries of real-world data. If such predictions can be obtained for training algorithms that also produce state-of-the art results, as tensor networks are starting to do, we anticipate this will continue to be an exciting program of research.

\bibliographystyle{unsrt}   
\bibliography{references}
\end{document}